# Augmenting learning in neuro-embodied systems through neurobiological first principles

**Alejandro Rodriguez-Garcia[1], Anindya Ghosh[1], Jie Mei[2,3,4], Srikanth Ramaswamy[1,5 *]**

[1]Neural Circuits Laboratory, Biosciences Institute, Faculty of Medical Sciences, Newcastle University NE2 4HH, UK
[2]International Research Center for Neurointelligence, University of Tokyo, Japan
[3]Department of Anatomy, University of Quebec in Trois-Rivieres, Canada
[4]IT:U Interdisciplinary Transformation University Austria, Austria
[5]Theoretical Sciences Visiting Program (TSVP), Okinawa Institute of Science and Technology Graduate University, Japan
* Correspondence: srikanth.ramaswamy@newcastle.ac.uk

## Abstract

Recent progress in artificial intelligence (AI) has been driven by insights from physics and neuroscience, particularly through the development of artificial neural networks (ANNs) capable of complex cognitive tasks such as vision and language processing. Despite these advances, they struggle with continual learning, adaptable knowledge transfer, robustness, and resource efficiency – capabilities that biological systems handle seamlessly. Specifically, neuromorphic systems and artificial neural networks often overlook two key biophysical properties of neural circuits: neuronal diversity and cell-specific neuromodulation. These mechanisms, essential for regulating dynamic learning across brain scales, allow neuromodulators to introduce degeneracy in biological neural networks, ensuring stability and adaptability under changing conditions. In this article, we summarize recent bioinspired models, learning rules, and architectures, and propose a framework for augmenting ANNs, which has the potential to bridge the gap between neuroscience and AI through neurobiological first principles. Our proposed dual-framework approach leverages spiking neural networks to emulate diverse spiking behaviors and dendritic compartmental dynamics, thereby simulating the morphological and functional diversity of neuronal computations. Finally, we outline how integrating these biophysical principles into task-driven spiking neural networks and neuromorphic systems provides scalable solutions for continual learning, adaptability, robustness, and resource-efficiency. Additionally, this approach will not only provide insights into how emergent behaviors arise in neural networks but also catalyze the development of more efficient, reliable, and intelligent neuromorphic systems and robotic agents.

## 1. Introduction

In recent years, AI has not only marked a significant evolution in technology but has also redefined the capabilities of machines, mirroring certain aspects of human intelligence with remarkable fidelity. At the forefront of this progress are advances in neural network architectures, which we refer to as artificial neural networks (ANNs) to distinguish them from the biological neural networks that form the basis of natural intelligence in living organisms. Among them, deep neural networks have proven especially transformative as their multi-layer architectures have demonstrated an impressive ability to learn from vast amounts of data[1–3].

Although ANNs are modeled to mimic biological neural networks, they still struggle with challenges that biological organisms master with ease, including adaptation, generalization, and learning across multiple timescales in an efficient and robust manner[4,5]. This disparity highlights a critical challenge: can we bridge the gap between artificial and biological neural networks to promote more flexible and adaptive learning in neuromorphic and robotic systems?

Bio-inspired ANN models are emerging as a promising solution to address this question. Designed to emulate the inherent flexibility and resilience of biological systems, these models aim to incorporate their learning mechanisms[5–7] in a neuromorphic energy-efficient setting[5,8–10]. Among different systems and properties of neurobiology, neuromodulatory systems are a good candidate for enhancing adaptive learning in artificial systems. Neuromodulators play a crucial role in supporting continual, sustainable, and cost-efficient learning in biological organisms by shaping behavior and function over multiple timescales – key attributes for adaptive learning and survival[11,12]. These chemical messengers are dispersed across a wide range of brain regions and neuronal populations simultaneously, influencing functional processes from synapses to neuronal firing patterns[12], much like neural operators in ANNs modulate responses based on specific inputs to optimize performance under varying conditions[13]. However, to fully understand how they drive adaptability from a biophysical perspective, it is essential to consider the diversity of neuronal populations that neuromodulators target.

In this context, it is essential to consider the brain's neuronal heterogeneity, i.e., diversity in the morphological, electrical, and physiological properties of neurons[14–18]. This diversity enables distinct neuronal populations to respond uniquely to neuromodulators, enriching the overall flexibility and adaptability of neural circuits[12,19–23]. This heterogeneity has often been overlooked in ANNs, as increasing the number of parameters required to model variability between different cell types increases computational costs[14]. However, recent studies using spiking neural networks (SNNs) have demonstrated that incorporating these heterogeneous populations adapts network computations[24,25], improves task performance and robustness[26–28], and facilitates multi-timescale learning[29,30]. Therefore, adding neuromodulators to heterogeneous SNNs could contextually, drive learning and adaptation, enhancing continual learning and enabling these systems to dynamically adjust to complex, changing environments across different timescales. This is important to maintain persistent stable dynamics in a network when faced with perturbation. Neuromodulators enable multiple pathways to exhibit similar macroscopic behaviors, mirroring degeneracy states in physical systems and thus introducing functional flexibility and resilience in biological neural networks[31,32].

In this article, we explore and describe an optimized framework, grounded in biophysics and dynamical systems, for applying the bio-inspired features of neuromodulation and neuron type specificity, extending a previous study on neuromodulation-inspired DNNs[11]. Hence, we first summarize recent developments in modeling ANNs with a focus on variations in neural network models, learning rules, and architectural designs. Building on this analysis, we generalize across solutions and propose a dual-framework approach that accommodates both biophysically plausible multi-compartmental models and SNNs of point neurons. By bridging the gap between neuromorphic spiking control systems and neuromodulatory-inspired learning strategies in large-scale SNNs, this framework aims to facilitate scalable and broadly applicable solutions to some of the most challenging issues in the field of AI, such as multi-timescale, multi-task, and few-shot learning, thus enabling robust and efficient models.

Moreover, it can provide insights into how complex biological processes can lead to the emergence of complex behaviors in biological systems.

## 2. From Isolated to Lifelong Learning

Learning can be defined as a dynamic process of restructuring a system to improve its task performance through experience[7,33,34]. Traditional AI approaches use machine and deep learning techniques to improve model performance on tasks through training, typically operating in supervised or unsupervised learning contexts[3]. Learning in these methods tends to be task-specific and trained on specific datasets (isolated learning[35]), demonstrating limited ability to process and retain contextual information and/or previously learned knowledge[35]. Additionally, they rely heavily on the training data and require extensive computational resources, limiting the capacity of the models to improve[8,9].

On the contrary, the dynamic nature of real-world scenarios requires biological organisms to develop and employ Lifelong Learning (LL), effectively managing ever-changing task demands and adapting to their environment to ensure survival. This capability, developed and refined through evolution, allows them to learn tasks sequentially and transfer and adapt task knowledge according to contextual changes – in a robust and efficient manner[5,7,34,35]. Inspired by these mechanisms, new learning paradigms such as online, multi-task, few-shot learning, and reinforcement learning are emerging[35].

### 2.1. Biological learning and neuromodulation

Evolution has enabled biological organisms to learn and adapt to uncertain and changing environments throughout their lives. Here we review the most important LL capabilities of biological systems and neural processes that support them:

***Learning continuously.*** Biological agents have evolved the ability to adapt to changing task demands throughout their lifetime to maintain their survival in natural environments. They integrate new information while retaining prior knowledge[6,34,35]. In AI systems, a similar process known as continual learning involves learning multiple tasks sequentially from continuous streams of information. However, this process is often limited by catastrophic forgetting, where new information is acquired at the cost of previously learned knowledge[36–38]. Since knowledge and tasks are provided incrementally over time, this term "continual learning" is often used interchangeably with LL, with many studies not making a strict distinction between the two[5,7,34].

Several biological processes in the brain serve as the basis for this capability. For example, at the macroscopic level, the states of sleep and wakefulness are controlled by a network of neuromodulatory nuclei that adjust their activity to regulate state transitions across the circadian cycle, thereby promoting dynamic memory storage for continuous learning[39,40]. At the mesoscopic level, meta-plasticity processes mediated by neuromodulators[41] and glial cells[42,43] further enhance continuous learning by selectively modulating synaptic connections. Similarly, dendritic spike-dependent plasticity also plays a critical role by selectively preserving and updating essential synaptic connections between neurons[44,45]. Additionally, neurogenesis is thought to support memory consolidation by generating new neurons, though its exact role in cognition is still uncertain[46,47]. Finally, at the sub-neuronal level, both ionotropic and metabotropic receptors are key in shaping both short- and long-term synaptic plasticity mechanisms, affecting continuous learning[48,49].

***Adaptive knowledge transfer.*** The dynamic nature of the environment drives living organisms to adapt rapidly. This encourages task-specific directional transfer of knowledge across tasks, either from an old task to a new task, or from a new task to an older task, to refine learned task representations. In addition, various species can detect changes in contingency and make inferences without identifying specific tasks[5]. As a result, life forms can learn quickly and efficiently from a few trials, a phenomenon known as few-shot learning[5,7,34].

**Box 1: Neuromodulation and physical degeneracy in neural networks.**

In statistical physics, degeneracy refers to a property of systems where different microscopic configurations yield the same macroscopic outcome[31,215]. This characteristic allows structurally distinct elements to perform equivalent functional outcomes or behaviors, contributing to the overall stability of the system. A classic example of this in thermodynamics is temperature. This macroscopic property can result from numerous combinations of different configurations of particle positions and velocities (microstates), all leading to the same thermal state[215]. Such intrinsic redundancy provides physical systems with significant adaptability and robustness against variations in internal or external conditions.

In biological systems, degeneracy is a fundamental property observed across different biosystems, from genetic structures and metabolic pathways to neural circuits, ensuring survival under uncertain conditions by providing adaptability and resilience[31,32]. In biological neural networks, degeneracy arises through neuromodulators that operate across spatial and temporal scales. Essentially, they enhance resilience and adaptability by enabling different neural pathways or mechanisms to produce similar functional outcomes and behavioral states, supporting flexible responses to environmental changes and robustness against perturbations[23,32]. Three main components illustrate how neuromodulators introduce neural degeneracy into neuronal systems:

- **Stability.** In both neural and physical systems, degeneracy promotes robustness by enabling diverse network configurations to reach a stable state. Neuromodulatory mechanisms support consistent neural circuit function by modulating signal-to-noise ratios[61,62], providing neuron-specific effects and making multi-timescale adjustments[12,187], which help regulate state transitions and maintain stability against perturbations[60].
- **Resilience.** Degeneracy enables physical systems to remain unaltered macroscopically, even when certain microscopic components malfunction, by compensating for their effects at a global scale[215]. Similarly, in neural systems, when one neuromodulatory pathway is impaired, other neuromodulatory systems can adapt to preserve essential functions to sustain survival like movement and cognition[32].
- **Adaptability.** Physical systems dynamically adjust to changes in state variables to maintain thermodynamic equilibrium[215]. Similarly, neuromodulators are essential for network adaptability, allowing neural systems to reconfigure and reach new equilibrium states in response to contextual triggers[19,21,32,56,57]. This degeneracy enables agents to adapt to changing tasks and supports continual learning throughout their lifespans[5,12,34,60].

Neuromodulatory nuclei modulate network behavior in response to changing contexts[11,12]. Specifically, histamine (HA), dopamine (DA), serotonin (5-HT), acetylcholine (ACh), and noradrenaline (NA) dynamically modulate the excitability of their neuronal targets, reshaping neuronal connectivity and altering brain and behavioral states to ensure survival in unpredictable environments[11,12,50–52]. For example, DA and 5-HT have been related to reward-driven behaviors[53,54], while ACh[55] and NA[56] have been related to arousal and attentional shifts to optimize behavior[57].

It is also important to note that the wiring rules of biological neural networks do not necessarily depend on experiences alone and can be established through evolution. These innate abilities in animals are part of a broader learning process that appears to be encoded in their genomes, a process that may be related to "supervised evolution"[58,59].

***Tolerance to noise.*** Real-world environments are characterized by uncertainty and noise. Nevertheless, biological organisms are largely robust to these conditions and respond optimally to their surrounding environment[60]. Particularly, the modulation of the signal-to-noise ratio by neuromodulators such as NA[56,61] and ACh[62] plays a crucial role in the robustness of biological systems to noise. Similarly, dendritic structures and homeostatic plasticity contribute to this tolerance by maintaining a high signal-to-noise ratio at potentiated synapses, ensuring that essential signals are discerned amidst the noise[63]. Furthermore, some studies suggest that neural cell type diversity contributes to resilience against pathological synchronization[64,65]. This appears to promote adaptability and robustness in learning, a recent finding supported by computational studies[24,25,27,28].

***Resource efficiency.*** Living organisms have limited resources that they must manage to achieve good performance in their tasks. This fundamental constraint forces them to become efficient by optimizing the use of their capabilities and energies to survive and thrive in a competitive environment[6,7,60]. Research has linked neural heterogeneity with efficient encoding of sensory systems[66]. Similarly, the multi-timescale effects of neuromodulatory systems, combined with the various cell-specific targets they influence, could also be responsible for the learning efficiency of natural neural systems[12,19,23].

Overall, neuromodulators introduce degeneracy in biological neural networks, thereby providing functional flexibility and resilience in biophysical systems[31,32] (see Box 1). These learning and adaptation mechanisms show how evolution has shaped the brain, not only to ensure survival but also to facilitate complex behavior and cognition in uncertain environments. Hence, understanding these processes and building upon biophysics and dynamical systems analysis offers valuable insights for designing more robust and adaptable AI models, potentially leading to systems that better mimic biological intelligence.

## 3. Neural network models

The way each neuron computes input signals in a neural network is fundamental, as it establishes the framework for the network model and its degree of biophysical realism. Therefore, ANN models can be classified in three main categories (see Table 1). These categories reflect varying degrees of adherence to biological principles, from the most abstract (top-down models) to the most detailed (bottom-up models).

| Network model | Description | Applications/Tasks |
|---|---|---|
| ***Abstract NNs*** $$y = f(\sum_{i=1}^{N} w_i x_i + b)$$ | $x_i$: neuron inputs<br>$w_i$: neuron weights<br>$b$: neuron bias<br>$f$: activation function (e.g., ReLU, tanh, SoftMax, sigmoid)<br>$y$: neuron output | Natural language processing[1].<br>Image classification and recognition[58,68,101].<br>Simulation and modeling[70].<br>Spatial navigation[69,71].<br>Decision making[72,164]. |
| ***Rate-based NNs*** $$\tau \frac{dr(t)}{dt} = -r(t) + \sum_{i=1}^{N} w_i x_i(t) + b$$ $$y(t) = f[r(t)]$$ | $x_i(t)$: neuron inputs<br>$w_i$: neuron weights<br>$b$: neuron bias<br>$r(t)$: neuronal activity<br>$\tau$: time constant<br>$f$: activation function (e.g., ReLU, tanh, SoftMax, sigmoid)<br>$y$: neuron output (firing rate) | Image classification[77,78].<br>Non-linear regression[77,78].<br>Multisensory integration[76].<br>Motor control[74–76].<br>Decision making[74,76,141]. |
| ***Spike-based NNs*** $$\frac{dv(t)}{dt} = a - bv(t) + \sum_{i=1}^{N} w_i x_i(t) + I_{ext}(t)$$ $$if\ v(t) \geq v_{th}: \begin{matrix} v(t) \leftarrow v_0, \\ y = \delta(t - t_0) \end{matrix}$$ | $x_i(t)$: neuron inputs<br>$w_i$: neuron weights<br>$I_{ext}(t)$: external current (noise)<br>$V(t)$: neuronal membrane potential<br>$\tau$: time constant<br>$a, b, v_{th}, v_0$: model parameters (LIF neuron)<br>$y$: neuron output (spike) | Natural language processing[93].<br>Image classification and recognition[28,92,96,97,137,188,200].<br>Simulation and modeling[83–85].<br>Cue association[91,92].<br>Multisensory integration[82].<br>Motor control[151,198,201].<br>Spatial navigation[152,154,155].<br>Decision making[91,149]. |

**Table 1: Overview of the three main types of neural network models, their mathematical definition, and applications.** Abstract NNs optimize for computational efficiency and simplicity by using a weighted sum of inputs plus a bias, thereby sacrificing biological accuracy. Rate-based NNs model neuronal activity through firing rates, incorporating time dynamics via differential equations. Spike-based NNs simulate the actual firing spikes of neurons, aiming to closely mimic the dynamic and temporal aspects of biological neural activity.

***Abstract neural networks.*** In these networks, neurons are modeled as a perceptron, where each neuron's output is a function of a weighted sum of its inputs, followed by a non-linear activation function[67]. The use of abstract representations in these neural networks reduces computational complexity, enabling processing of large volumes of data and complex tasks at low computational costs[2,3]. Hence, they have been widely employed in deep learning applications such as computer vision[58,68], natural language processing[1], and navigation and decision-making tasks[69–72].

***Rate-based neural networks.*** From a theoretical neuroscience standpoint, the firing rate of a biological neuron represents its average activity measured by the number of spikes over a time interval as a continuous variable[73]. Inspired by the computation of these, rate-based neural networks define their activity through time-dependent firing rates, thereby accounting for the temporal representations of data and signal processing. This makes them useful for studying phenomena such as plasticity and learning at a neural level. As such, they are used in tasks such as perceptual decision-making and motor control[74–76], as well as in image classification tasks that exploit bio-inspired insights into backpropagation signals[77,78].

***Spike-based neural networks.*** Initially developed in computational neuroscience to depict single-cell level activities, spike-based NNs, also known as spiking neural networks (SNNs), simulate the precise neuronal spikes with the aim of simulating the temporal aspects of neuronal activities[73]. However, the biological complexity of neuronal spikes leads to a wide array of models, each varying in complexity and degree of biological inspiration. Therefore, these models are broadly classified into conductance-based models, which explore the biological mechanisms of neuronal activity, and phenomenological models, which describe behavior based on observable phenomena[73,79,80]:

- *Conductance-based models.* Grounded in the biophysical Hodgkin and Huxley (HH) model, conductance-based models accurately describe how membrane potentials behave at a sub-neuronal level[73,81]. Additionally, given their biophysical fidelity, these models incorporate dendritic structures through multiple compartments[82–85], and typically run on simulation environments like NEURON[86,87], or Brian 2[88]. In this last simulator, the Dendrify environment has recently gained importance as a versatile framework that simplifies the incorporation of dendritic structures into SNNs[89].
- *Phenomenological models.* The intrinsic complexity of the HH model has prompted the development of simplified variants that generate spiking behaviors through dynamical systems and bifurcation theory[73,79,80]. The simplest case, the leaky integrate-and-fire (LIF) neuron, captures the basic firing mechanism of neurons[79,90]. Its simplicity and scalability make it popular in SNNs[91–93] and neuromorphic hardware[8,9,94,95]. However, this simplicity also limits its ability to represent specific neuronal activities, leading to the development of numerous variants that incorporate different biological features[79,96–98].

### 3.1. Neuronal model tradeoffs

Effective network design requires a careful balance between model complexity and biophysical fidelity, tailored to the specific demands of the intended tasks. To enhance the continual learning capabilities of ANNs through neuromodulators, it is essential to account for neuromodulatory effects across spatial scales, including their long-term modulation of spiking behavior. The generalized leaky integrate-and-fire (GLIF)98 and the Izhikevich (IZ)99 models stand out as optimal choices due to their balance between complexity and bioinspiration (Figure 1).

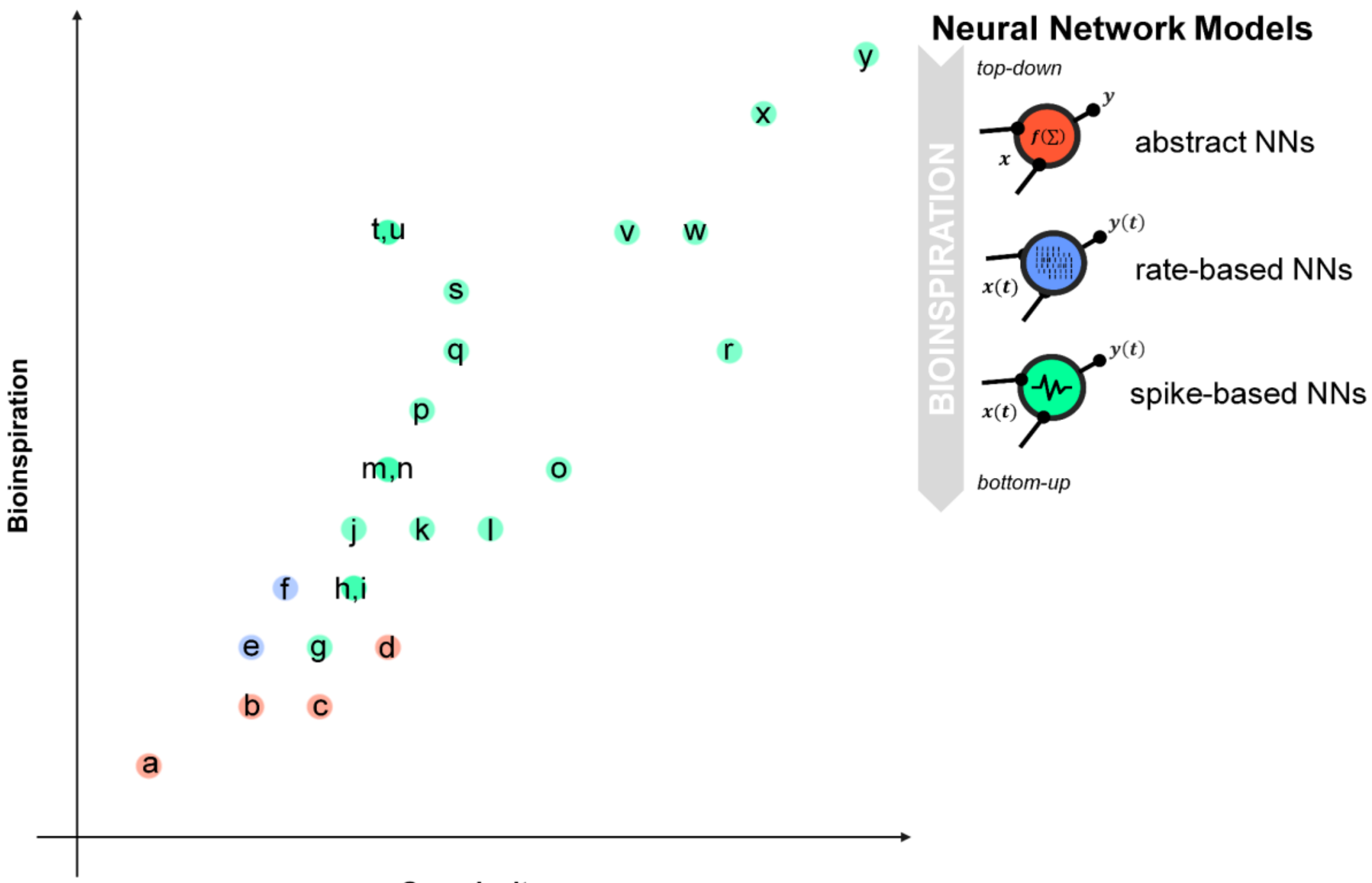

**Figure 1: Bio-inspired neurons in ANNs.** *Left:* Scatter plot of the bioinspiration and complexity trade-off in neuron modeling approaches. The x-axis quantifies the degree of complexity of the neuron model and the y-axis the degree of bioinspiration (both normalized). These metrics have been calculated based on categorization and comparison of the models in the literature. The neuron models can be classified according to their network type. *Abstract Neural Networks* (red): **(a)** perceptron neuron[67,202], **(b)** gated recurrent unit (GRU)[203], **(c)** long short-term memory (LSTM) network[204], **(d)** evolvable neural unit (ENU)[70]. *Rate-based neural networks* (blue): **(e)** rate-based neuron[73], **(f)** compartmental rate neurons[77,78]. *Spike-based Neural Networks* (green): **(g)** leaky integrate-and-fire (LIF) neuron[73,90], **(h)** adaptative leaky integrate-and-fire (ALIF) neuron[91,98], **(i)** integrate-and-fire with adaptation (IFwA) neuron[79], **(j)** quadratic integrate-and-fire (QIF) neuron[205], **(k)** leaky integrate-and-modulate (LIM) neuron[97], **(l)** exponential integrate-and-fire (EIF) neuron[206], **(m)** resonate-and-fire (RF) neuron[207], **(n)** integrate-and-fire-or-burst (IFoB)[208], **(o)** FitzHugh-Nagumo model[209], **(p)** parametric leaky integrate-and-fire (PLIF) neuron[96], **(q)** Spike Response Model (SRM)[210], **(r)** Morris-Lecar model[211], **(s)** adaptive exponential integrate-and-fire (AdEx) neuron[212], **(t)** Izhikevich (IZ) neuron[99], **(u)** generalized leaky integrate-and-fire (GLIF) neuron[98], **(v)** Hindmarsh–Rose neuron[213], **(w)** Wilson neuron[214], **(x)** Hodgkin–Huxley (HH) neuron[81] and **(y)** biophysical multi-compartmental models[86,88]. Right: Representation of the bioinspiration continuum in NN models. Top-down approaches involve abstract NNs that focus on the neuron's inputs, applying a function to their weighted sum to address non-linearities, with both inputs and outputs being real numbers. On the other hand, bottom-up approaches as spike-based NNs consider biophysical neurons from computational neuroscience capturing spiking activity. In this scenario, inputs and outputs are represented by neuronal spikes. Bridging these two approaches are the rate based NNs, which model biological neuron computation in terms of the neuron firing rates.

These models support various types of spiking patterns through parameter adjustments, making them ideal for simulating diverse neuronal populations without excessive computational load (see Box 2). On the other hand, capturing morphological heterogeneity, specifically dendritic dynamics, requires multi-compartmental biophysical models, which demand substantial computational capacity. In this context, the Dendrify framework emerges as an efficient option to model dendritic compartments[89].

**Box 2: The Izhikevich model and the GLIF neuron simulate different spiking behaviors.**

The Izhikevich (IZ) model is a phenomenological model rooted in bifurcation theory, designed to simplify conductance-based models using a small number of parameters while preserving various spiking behaviors[79,99]. Its dynamics are represented through a set of two coupled differential equations:

$$\frac{dv(t)}{dt} = 0.04 \cdot v(t)^2 + 5 \cdot v(t) + 140 - u(t) + I_{ext}(t),$$

$$\frac{du(t)}{dt} = a \cdot \big(b \cdot v(t) - u(t)\big).$$

With an auxiliary after-spike reset:

$$if\ v(t) \geq\ +30\ mV \Rightarrow \begin{cases} v(t) \leftarrow c \\ u(t) \leftarrow u(t) + d \\ y(t)\ =\ \delta(t - t_s) \end{cases}.$$

Here, $I_{ext}(t)$ is the external current, $v(t)$ is the membrane potential, $u(t)$ represents the membrane recovery variable, $y(t)$ the spike record as the output of the neuron; and $a$, $b$, $c$ and $d$ are certain parameters. Note that in this model, the spike threshold is set at $+30\ mV$, and the scale is chosen in a way that the time has $ms$ units[99]. This model has demonstrated the simulation of the spiking behaviors of most cortical neuron types[79], as well as polychronous patterns consistent with in vivo recordings during working memory tasks[148,216] and shown the distal reward problem when integrated with DA signaling[146] in simulations of large-scale networks. Therefore, it is a compelling option for simulating diverse neuron populations across extensive neural networks[24,25]. However, the quadratic terms in the differential equations make it computationally more expensive than the LIF model for neuromorphic hardware[217].

On the other hand, the GLIF neuron enhances the basic LIF model by incorporating linear differential equations along with specific after-spike rules that allow to simulate the spiking behavior of different cell types, making it an interesting option for neuromorphic hardware[184]. Mathematically, the GLIF neuron is described as[98]:

$$\frac{dv(t)}{dt} = \frac{1}{C} \cdot \big(I_{ext}(t) + \textstyle\sum_j I_j(t) - \frac{1}{R} \cdot (v(t) - E_L)\big),$$

$$\frac{d\theta_s(t)}{dt} = -b_s \cdot \theta_s(t),$$

$$\frac{dI_j(t)}{dt} = -k_j \cdot I_j(t);\ j = 1{,}2,$$

$$\frac{d\theta_v(t)}{dt} = a_v \cdot (V(t) - E_L) - b_v \cdot \theta_v(t).$$

Here, $I_{ext}(t)$ is the external current, $v(t)$ is the membrane potential, $\theta_s(t)$ is the spike-dependent component of the threshold, $I_{j=1,2}(t)$ two after-spike currents and $\theta_v(t)$ the membrane potential-dependent component of the threshold. Additionally, $C$ is the neuron's capacitance, $R$ the membrane resistance, $E_L$ the resting potential, $a_v$ the membrane coupling to the threshold, and $\frac{1}{b_s}, \frac{1}{b_v}, \frac{1}{k_j}$ are the time constants of the spike and voltage dependencies of the threshold, and the time constants for the after-spike currents, respectively. The after-spike rule is posited as follows:

$$if\ v(t) \geq\ \theta_v(t) + \theta_s(t) + \theta_\infty \Rightarrow \begin{cases} v(t)\ \leftarrow E_L + f_v\ \cdot (V(t) - E_L) - \delta v \\ \theta_s(t) \leftarrow \theta_s(t) + \delta\theta_s \\ I_j(t) \leftarrow f_j \cdot \big(I_j(t)\big) + \delta I_j \\ \theta_v(t) \leftarrow \theta_v(t) \\ y(t)\ =\ \delta(t - t_s) \end{cases}.$$

Where $f_v$ and $\delta v$ denote the intercept and slope of the linear relationship between the voltage before and after a spike, $\delta\theta_s$ is the threshold post-spiking increase, $f_j$ and $\delta I_j$ represent the fraction and amplitude of the after-spike currents; and $y(t)$ is the spike record as the output of the neuron. This model has also been demonstrated to reproduce the spiking behaviors of the Allen Cell Types Database[98], thereby being an interesting option to simulate the spiking activity of diverse cell populations.

## 4. Task-driven learning strategies

The choice of an appropriate learning algorithm is strongly related to the desired model architecture and the level of biophysical fidelity needed. In this way, each algorithm presents distinct advantages and limitations, which must be carefully considered in relation to the specific task at hand. We analyzed these algorithms across a spectrum of cognitive and AI tasks (Figure 2), revealing nuanced insights into their applicability and efficacy.

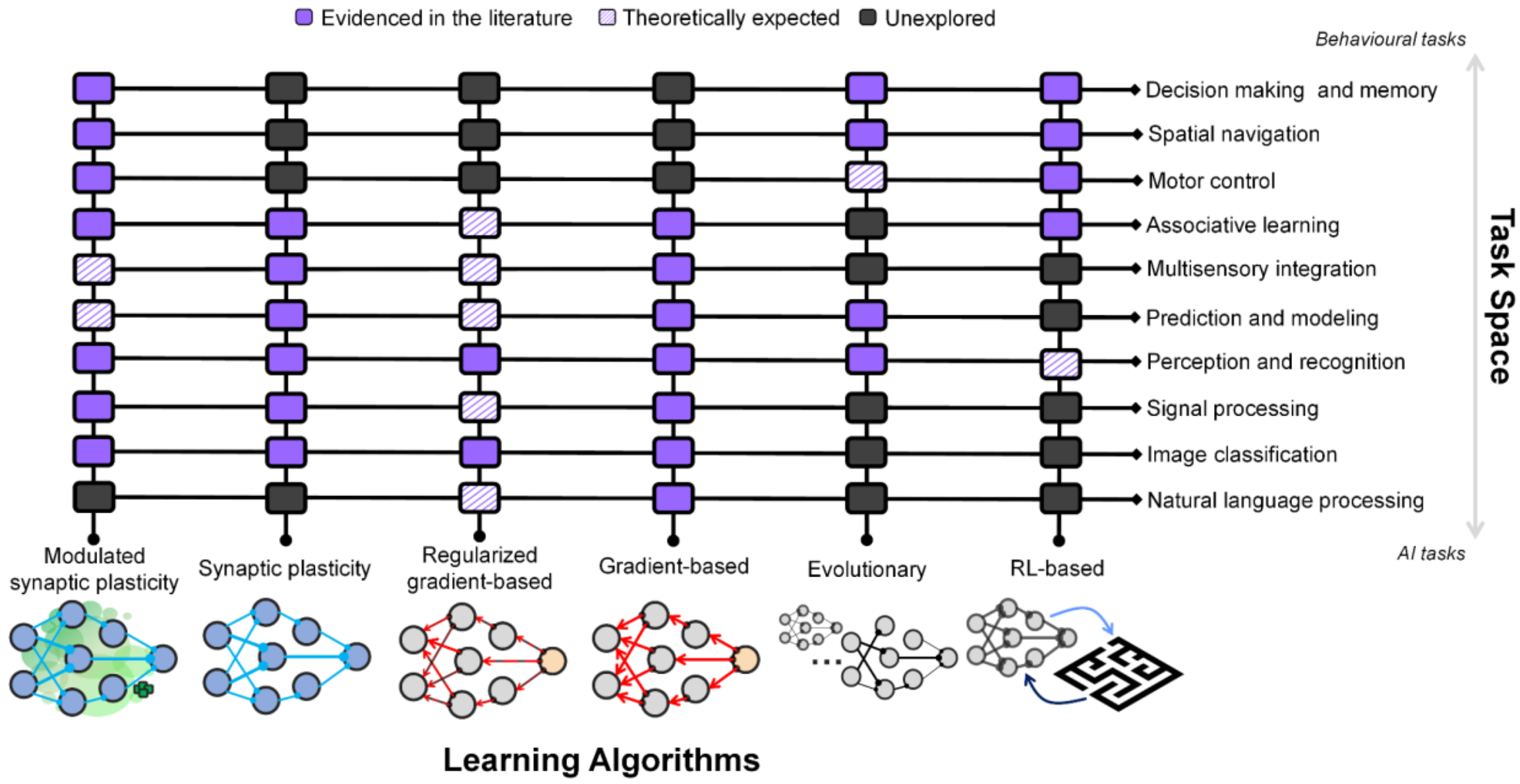

**Figure 2: Taxonomy of learning algorithms with respect to the task space.** The vertical axis represents a task space from AI tasks to behavioral ones, and the horizontal axis the types of learning algorithms, each illustrated with a schematic representation of neural connections. Additionally, we classified the tasks into those for which models have been identified in the literature (purple filling), those that are anticipated to be found but have not yet been considered or discovered in the literature (purple stripes), and those that have not been explored (gray filling). The diagram shows that gradient-based methods are basically restricted to AI-type tasks, even though they can be integrated into behavioral tasks through Deep RL frameworks using RL-based algorithms. Similarly, evolutionary algorithms are considered for behavioral tasks and modeling as they are more robust than gradient-based approaches when the optimization is complex. Finally, modulated synaptic plasticity methods emerge as the most versatile option. Their adaptability across various task types is due to the combination of a local learning rule and a context modulatory signal that enhances learning.

***Gradient-based algorithms.*** These algorithms optimize a loss function using gradient descent, thereby updating model parameters to improve performance[2,3]. Their foundation lies in the Backpropagation algorithm, which uses the chain rule to propagate errors backward through abstract and rate-based NNs[100]. This algorithm effectively addresses the credit assignment problem in ANNs, and its high performance makes it widely used for tasks like natural language processing[1,93], image and speech recognition[68,101,102], and signal processing[103,104]. However, its biological plausibility is often questioned, though ongoing efforts aim to interpret and validate its mechanisms in biological terms[69,105–108].

To enhance the applicability of Backpropagation across different NN models, numerous variants have been developed. For instance, it has been adapted as BPTT to handle temporal dynamics in RNNs[105], while alternative gradient-based methods, such as FORCE[109] and full-FORCE[110], train RNNs for behavioral tasks by adjusting only the output weights, treating the recurrent network as an input-perturbed reservoir. A recent bio-inspired approach, Feedback Alignment (FA), adapts BP as a local learning rule aligned with neural plasticity, achieving notable results in image classification[111] and signal processing tasks[112]. Furthermore, BP has been extended to SNNs by applying surrogate gradients to the non-differentiable spikes[113,114]. Particularly, eligibility propagation (e-prop) utilizes this technique along eligibility traces to backpropagate signals in an efficient and bio-inspired manner, achieving high performances in supervised and reward-based tasks through Deep RL frameworks[91,115].

Overall, gradient-based methods excel in isolated learning paradigms, efficiently handling deep layers and numerous parameters inherent to abstract neural networks, essential for state-of-the-art AI tasks[2]. However, their susceptibility to catastrophic forgetting necessitates regularization methods to enhance their robustness and applicability in LL scenarios[88].

***Regularized gradient-based algorithms.*** Gradient-based algorithms can incorporate regularization techniques like Dropout to prevent overfitting and enhance performance in abstract NNs[116]. Recently these techniques have evolved to enable continual learning capabilities across different tasks by preventing the synaptic weights from overwriting, as happens with orthogonal weight modification (OWM)[117], memory aware synapses (MAS)[118], Sliced Cramér Preservation (SCP)[119], synaptic intelligence (SI)[120] or elastic weight consolidation (EWC)[121]. Particularly, these last two methods draw inspiration from dendritic synaptic plasticity and have been proven to be similar to what biological neurons use, involving NMDA-mediated plasticity and the grouping of closely related synapses[45,122–125]. Despite their proven capabilities[126], their focus on image classification and perception has left them underexplored in a broader range of real-world applications.

***Synaptic plasticity algorithms.*** Grounded in neuronal learning mechanisms and the Hebbian principle[127], synaptic plasticity methods are characteristic of spike-based NNs[47], though some variants are also found in abstract NNs[1,128]. In the SNN domain, the spike-timing-dependent plasticity (STDP) algorithm stands out as a *stable* and bio-inspired approach that emphasizes the exact timing of spikes between pre- and post-synaptic neurons[129–131]. Hence, many studies have implemented this rule in unsupervised contexts, solving tasks ranging from time series prediction[132] to pattern[133] and image classification[134]. However, for greater bio-plausibility, some research opts for a triplet-based variant, which considers one pre-synaptic spike and two post-synaptic spikes, accounting for higher spatiotemporal correlations [92,135]. Additionally, newer methodologies, like Burstprop and Burst Ensemble Multiplexing (BEM), leverage synaptic bursting activity, backpropagating signals in rate-based neurons with dendritic compartments, and supporting supervised learning in perceptual and image recognition tasks[77,136,137].

***Modulated synaptic plasticity algorithms.*** Despite the inherent bio-inspiration of synaptic plasticity algorithms, they become significantly more powerful when integrated with modulatory signals, which enhance their learning capabilities, thus transforming these methods into neo-Hebbian methods[138,139]. Methods like gated Hebbian rules[140], rarely correlating Hebbian plasticity (RCHP)[131,141], and the exploratory Hebbian (EH) algorithm[75,142], modulate synaptic plasticity with reward signals or context-aware mechanisms, achieving good performance in image classification, cue-association, and motor control tasks.

Nevertheless, the STDP rule is particularly interesting for integrating these modulatory signals, as neuromodulatory signals regulate its window of plasticity induction[41,143,144]. In this context, the rule is known as reward-modulated STDP (R-STDP), which adjusts the plasticity window based on a varying reward signal[145]. The most direct use of this signal acts as a form of dopamine (DA) modulation influencing synapses[146–148], linking it to the reinforcement scenarios as in TD-STDP[149] or actor-critic architectures[150–152].

Additionally, some studies extend these modulated STDP algorithms further, attempting to consider the effects of various neuromodulator signals, including co-release of DA and ACh[41,153,154], as well as DA and 5-HT[155]. These methods are particularly intriguing as they offer ways to incorporate ANNs into reinforcement or unsupervised learning scenarios, reflecting the adaptive nature of biological learning processes. Hence, modulated STDP rules stand out as versatile methods that can be easily integrated into reinforcement paradigms[11,50]. Furthermore, these rules can integrate adaptive learning through context-aware signals that have a biological foundation in neuromodulators.

***Evolutionary algorithms.*** Inspired by Darwinian evolution, evolutionary algorithms were created in computer science to solve complex nonlinear optimization problems by mimicking the iterative process of biological evolution, focusing on the survival of the fittest solutions[156,157]. Their robustness and adaptability to various NN models make them better-suited to behavioral tasks, excelling where gradient-based methods struggle due to the complexity of optimizations[158,159]. Additionally, they have been employed to model specific brain behaviors, offering insights into neural mechanisms and contributing to our understanding of cognitive functions[70,83].

***RL-based algorithms.*** Inspired by behavioral psychology, RL-based techniques enable learning through the interaction of an agent with its environment. The agent interacts with its environment by taking actions that alter states and yield rewards, enabling it to improve its policy and optimize its behavior.[160]. Traditionally used with abstract models, RL has been significantly enhanced by integrating with ANNs, leading to Deep RL. This advancement combines RL with deep learning's powerful function approximation, creating LL paradigms for reinforcement agents[6,60,160].

In the RL context, the Temporal Difference (TD) learning rule is particularly crucial due to its association with DA as the global learning signal for reward predictions[161,162]. Expanding on this, additional neuromodulators have been tied to specific RL parameters: 5-HT regulates the balance between short-term and long-term reward predictions, NA adjusts the balance between exploration and exploitation in the policy, and ACh influences the learning rate of the agents[11,12,50,163]. This integration enhances decision-making tasks and improves the adaptability and efficacy of continual learning agents, thereby providing an excellent framework for integrating neuromodulatory signals to ANNs[50,60].

These methods can be integrated through meta-learning processes, enhancing the versatility of models to solve tasks for which they are not optimal. Therefore, different learning rules can be employed for different learning

**Box 3: Neuromodulatory-inspired learning.**

Neuromodulation is intricately linked to meta-learning. This concept is often described as "learning to learn" and it is defined as the system’s capability to adapt its learning process. Meta-learning involves integrating multiple learning loops within the same system, enabling it to tackle tasks across various timescales[35]. Therefore, meta-learning procedures enable the use of more than one learning algorithm to learn the different learning loops. For instance, an inner loop might be governed by synaptic plasticity and the outer loop by gradient descent, or the outer loops can correspond to evolutionary or RL-based algorithms[92,158,164]. In this context, neuromodulators can be responsible for different learning loops influencing plasticity and network tuning. In SNNs, neuromodulatory systems are directly incorporated through modulated or third-factor synaptic plasticity rules, introducing a contextually driven influence at the circuit connectivity. They have been also integrated within reinforcement learning paradigms, explaining single and multi-neuromodulatory dynamics in navigation and motor control tasks[11,50,141,151,153–155]. In abstract and rate-based NNs, the generalizability of this concept extends to govern hyperparameter learning, tuning network optimization in specific tasks[165,218,219].

Similarly, neuromodulatory-inspired learning is believed to regulate transitions between different network states, such as sleep-wake cycles or varying levels of arousal, which subsequently adjust the model's response to external stimuli and support continual learning[25,40]. In this sense, neuromodulators in biological systems can be viewed as analogs to neural operators in artificial systems. Neural operators provide high-fidelity mappings across continuous domains, acting as adaptive mechanisms that maintain system performance under shifting conditions[13]. Just as neuromodulators in biological networks dynamically adjust neuronal processing based on context, neural operators enable similar adaptability in artificial systems.

loops, relevant to multi-task learning[92,164]. Additionally, neuromodulation can play a crucial role in controlling or participating in these processes (see Box 3), enriching the models' ability to adapt and perform across diverse applications[11,165].

## 5. Bio-inspired architectures

ANN architectures fundamentally rely on two primary structures differentiated by the flow of information: feedforward neural networks (FFNNs), where information flows linearly from inputs to outputs, and recurrent neural networks (RNNs), where information is also looped back into the network through recurrent connections among neurons[73]. With the development of deep learning, the inclusion of deep-hidden layers was initially pursued to handle more complex tasks[2,3]. However, this approach appears to have reached its limits, prompting a shift towards integrating bio-inspired features that mimic LL capabilities found in biological systems[6], thereby generating new bio-inspired architectures (Figure 3).

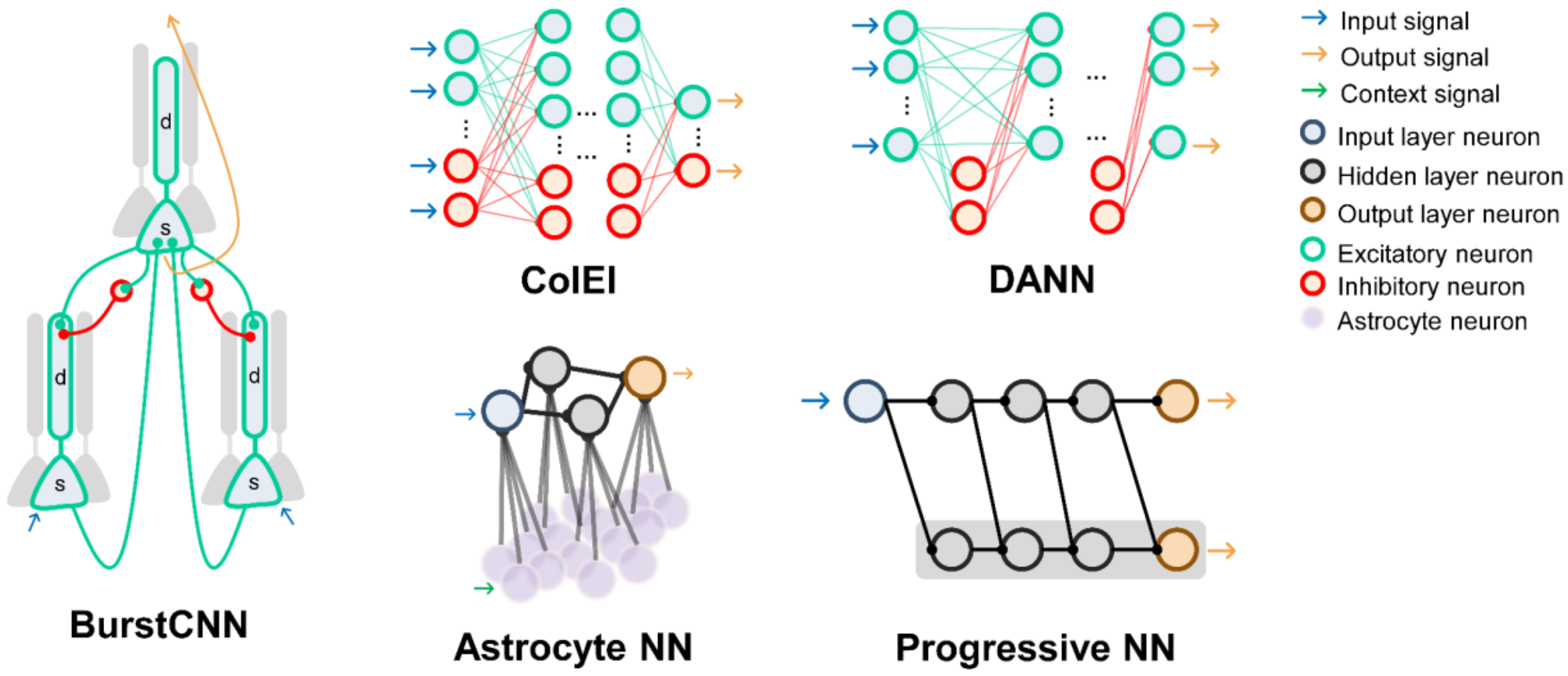

**Figure 3: Bio-inspired architectures of ANNs.** Emerging architectures inspired by biological principles are making their way into ANNs, particularly those that incorporate both excitatory and inhibitory neuron populations, such as Column Excitation-Inhibition (ColEI)[179] and Dale's ANN (DANN)[178,179]. These systems utilize Dale's principle to define neuronal populations with specific synaptic weight constraints (positive weights for excitatory neurons and negative weights for inhibitory neurons). Dendritic-like architectures such as Bursting Cortico-Cortical Networks (BurstCNN)[77,78,137], model pyramidal neurons with excitatory compartments - the soma and the apical dendrite, in this case - and interneurons as one inhibitory compartment, thereby offering a biologically plausible method of backpropagating signals in the brain. Astrocyte neural networks[97,181,182] expand beyond traditional neural populations to include models of glial cells, providing a deeper understanding of neural systems by integrating the role of astrocytes in memory consolidation and attention. Progressive neural networks[173], inspired by hippocampal neurogenesis, address catastrophic forgetting in ANNs by generating task-specific sub-modules.

### 5.1. Dendritic compartmental dynamics

Dendritic architectures emerge in the context of biophysical models, with dendritic trees being modeled as electrical circuits consisting of multiple compartments[63]. However, these models are complex and computationally demanding, prompting the development of sophisticated architectures designed to emulate their capabilities within abstract NNs of point neurons[166]. This has led to the incorporation of dendritic-like processing in network structures, which aims to capture the input properties of the neuron with a more efficient description[167]. Here, dendritic computations can be effectively represented using two- and three-layered feedforward network models. In this approach, inputs are integrated nonlinearly through an activation function, and then signals are further integrated to the neuron. These activation functions are often sigmoid functions[168,169], but other approaches utilize sub- and supra-linear functions to more accurately represent dendritic behaviors[85].

Currently, dendrites, as they contribute to continuous learning and adaptation processes in the brain, have become an essential feature for supporting continual learning in ANNs[45,122]. Particularly, some approaches use context-driven dendritic layers in abstract NNs[170] or NMDA-driven modulation in spike-based NNs[171] for multi-task learning, while others employ rate- and spike-based NNs with burst-dependent plasticity learning algorithms, giving rise to the so-called Bursting Cortico-Cortical Networks (BurstCCN), thereby solving the credit assignment problem by back-propagating signals in the dendrites[77,78,137]. Additionally, dendritic trees enhance ANN efficiency, allowing fewer neurons to handle complex tasks. For instance, they have been demonstrated to solve the XOR problem with a two-compartment single neuron model, which typically requires multiple layers of point neurons[172].

**5.2. Network modularity and progressivity**

Dynamic architectures - inspired by neurogenesis in the hippocampus - adapt by adding new neuronal resources while training for new tasks, to preserve knowledge in response to new information[46]. Similarly, modular architectures directly utilize parallel sub-modules to adaptively address specific tasks. These networks are aligned with the brain's functional connectivity, offering a solution for multi-task learning and knowledge transfer[5].

Progressive networks exemplify both strategies by incrementally expanding their abstract NN architecture with new sub-networks dedicated to each new task while preserving a collection of pre-trained modules for each task already learned[173,174]. However, this design, although it facilitates the simultaneous learning of multiple tasks, results in escalating complexity as the number of tasks increases.

**5.3. Population-Based architectures**

Most bio-inspired architectures have primarily focused on incorporating both excitatory and inhibitory neuron populations. They are based on Dale's law, which posits that a neuron releases the same neurotransmitters at all its synapses, influencing other neurons in a consistent manner – either by exciting or inhibiting them[175]. However, recent research has refined Dale's law, showing that neurons can release different combinations of neurotransmitters and that the effect on the target neuron varies depending on its specific receptors[176,177]. Despite this, numerous computational models use this principle to consider different populations of excitatory and inhibitory neurons. They assume that excitatory presynaptic neurons make purely excitatory (positive weights) connections with postsynaptic neurons and inhibitory ones make purely inhibitory (negative weights) connections[76,142,178].

The presence of these two neuronal types has led to the development of new neural network architectures, ranging from RNNs to Deep Learning structures, adding a biological constraint to any type of NN model. Some use an RNN with populations of excitatory and inhibitory neurons in a 4:1 ratio[74,76,91,146]. Others incorporate this biological constraint into the feed-forward connections of DNNs or CNNs, resulting in an architecture called column excitation-inhibition (ColEI)[179]. However, since this structure usually impairs learning, Dale's principle is typically left out of abstract NNs[178–180]. This issue led to the development of Dale's ANN (DANN) architecture, inspired by feedforward inhibitory interneurons in the brain. Here, every layer of the network is composed of either excitatory or inhibitory units, following the biological proportion. Furthermore, only excitatory units can project between layers, while inhibition is managed by activation rules that allow effective modulation between both populations. Thus, inhibition is not constrained by the sign of the synaptic weight and DANNs does not sacrifice learning performance[178,179].

However, simply incorporating excitatory and inhibitory populations is not sufficient to capture the complexity of biological systems. To further enhance the bio-inspiration of ANNs, it is crucial to integrate diverse spiking behaviors in SNNs (see Box 2), which improve their adaptive capabilities[24,25,28].

**5.4. Astrocyte networks and transformers**

Recent models extend beyond neuronal networks by incorporating astrocytes, recognizing their crucial role in cognitive functions[42]. Therefore, new perspectives are seeking bio-inspired architectures that incorporate an astrocyte NN guided by contextual triggers to modulate neurons in ANNs[181]. In this direction, neuron-astrocyte liquid state machine (NALSM) architectures have also been proposed to account for the modulation of synapses by astrocytes, achieving performance levels like those of multi-layer SNNs trained with backpropagation[97].

Another study links the astrocyte network within an abstract NN by constructing a neuron-astrocyte model that replicates the functionality of a Transformer[182]. Transformers represent a complex network architecture utilized in natural language processing, distinguished by their self-attention mechanism[183]. This feature enables them to capture long-range dependencies between words in a sentence efficiently, eliminating the need for maintaining a hidden state over extended periods, as required in recurrent models. However, these architectures are heavily reliant on big data on have high energy usage, contrary to learning in biological systems[182].

## 6. Neuromodulatory-aware spiking networks

The field of ANNs is constantly evolving, witnessing emerging learning rules, architectural designs, and morphological features inspired by biological systems to improve continual learning, adaptability, robustness and resource efficiency in ANNs[5–7]. To further address these challenges, we propose a dual-framework approach that combines biologically informed dendritic models and task-driven SNNs of heterogeneous neurons to facilitate improved performance and efficiency in ANNs (Figure 4). Our approach balances biophysical fidelity and computational efficiency by the simulation of neuronal activity using the IZ[24,25,99] and GLIF[98,184] models and imposing constraints to the synaptic weight to define excitatory and inhibitory populations. This allows to account for the spiking heterogeneity that characterizes neural systems[15–18] and brings the opportunity to account neuromodulatory cell-specific effects to the network[12,21–23]. This approach has the potential to incorporate neuron specificity and neuron-specific neuromodulatory signals, while maintaining the highest level of simplicity possible, making it reasonably cost-efficient even in large-scale SNNs.

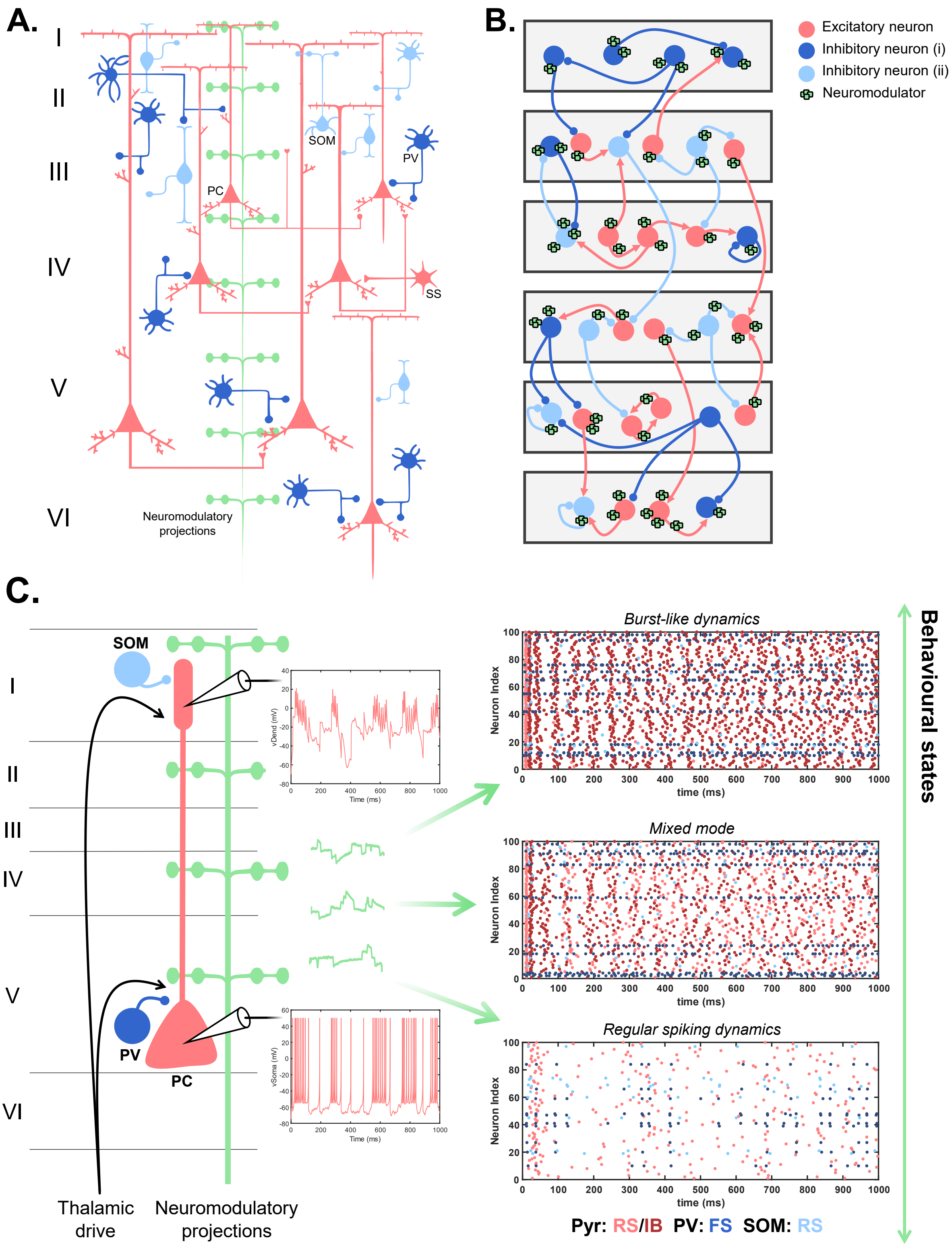

**Figure 4. Neuromodulatory-aware heterogeneous SNN framework for continual learning.** **(A)** Cortical microcircuit representation. (A) A schematic of a biological neural network, illustrating a cortical microcircuit with the main cell types in the neocortex: pyramidal neurons (PC), spiny-stellate cells (SS), parvalbumin-positive (PV), and somatostatin-positive (SOM) interneurons. Neuromodulatory projections regulate neuronal activity and shape network responses to contextual changes. To model this complexity, biophysical multicompartmental models are often used, capturing dendritic processing and synaptic integration. However, these models present scalability challenges due to their high computational cost. **(B)** To address these limitations, we introduce a spiking neural network (SNN) model optimized for AI tasks while retaining key features of neuronal

heterogeneity and neuromodulatory dynamics. This approach incorporates excitatory and inhibitory neurons with distinct spiking properties, dynamically influenced by neuromodulation to enhance task-specific learning. **(C)** Example of the neuromodulatory-driven adaptive dynamics. This two-compartment neuronal model illustrates how neuromodulatory signals enable real-time transitions between different network dynamics by modulating the connectivity between the apical and somatic compartments and adjusting the gain of pyramidal neurons. These changes allow neurons to switch between regular spiking and bursting regimes, facilitating dynamic reconfigurations in network activity. Ultimately, this mechanism enhances the network's ability to flexibly respond to changing input conditions, improving adaptability in both AI and neurobiological systems. Different neuronal types exhibit diverse spiking patterns: regular spiking (RS), bursting (IB), fast spiking (FS).

### 6.1. Neural heterogeneity and lifelong learning

The biological brain is characterized by its cell type diversity, featuring a wide variety of neurons with different morphologies and spiking behaviors[14–18,79], which influence how they process temporal information. In contrast, most ANNs use uniform sets of neurons to process information, restricting their learning to plasticity dynamics. Despite being traditionally overlooked[14], recent computational studies suggest that the incorporation of neural heterogeneity contributes to LL by improving task performance[28], promoting stable and robust learning[26–28], facilitating multi-timescale learning[29,30], and efficiently adapting computations in SNNs[24,25].

Practically, in SNNs, neural heterogeneity can be mainly introduced either morphologically, by defining distinct dendritic compartments for different neuron types, or functionally, by simulating different spiking behaviors. Morphological heterogeneity has been explored through dendritic architectures by defining cell types, such as pyramidal neurons, interneurons, or basket cells, with different dendritic compartments[30,85,137]. Particularly, such efforts have helped mitigate catastrophic forgetting, solve the credit assignment problem, and handle tasks efficiently[45,122,172].

Functionally, neuronal diversity can be represented by distinguishing excitatory and inhibitory populations through synaptic constraints that adhere to Dale's law. However, in spike-based models, this can be further considered by tuning the neuron's model parameters (for a full description, see Box 2) to simulate different spiking behaviors such as regular spiking, fast spiking, bursting, or threshold variability[79,98]. Notably, the use of heterogeneous spiking patterns has been particularly explored in spiking control systems[185,186], where defined connected networks of different spiking motifs are able to exhibit circuit rhythms computationally, though they face scalability limitations.

Our approach aims to exhibit many different spiking behaviors while being scalable, by relying on the IZ and GLIF models (see Box 2). In this context, introducing cell-specific spiking behaviors can potentially adapt computations in SNNs. Particularly, theoretical studies in this matter using a mean-field model of coupled IZ neurons have proven that heterogeneity in inhibitory interneurons gates signaling propagation while heterogeneity in excitatory neurons improves the capacity of neurons for decoding tasks[24]. Another study indicates that neuronal time-scale heterogeneity leads to robust learning and a performance improvement of SNNs with LIF neurons for an image classification task[28]. Similarly, a recent study extends these findings to bursting parameters using evolutionary algorithms to solve spatiotemporal tasks[26]. Ultimately, the current focus on learning the synaptic connections in ANNs seems quite restrictive. Therefore, incorporating neural heterogeneity is essential to provide more degrees of freedom to optimize neural networks without increasing the number of neurons, thereby enhancing learning and stability.

### 6.2. Neuromodulation for continual learning

Neuromodulatory systems provide a multi-layered contextual-guided control over cognition and behavior in nervous systems. They operate on distinct temporal scales and influence both spiking behavior at the neuronal level and global network plasticity properties at the circuitry level, thereby reconfiguring network dynamics at a systems level[12,19–23] and introducing network degeneracy[32] (see Box 1). Particularly, spiking control systems have leveraged neuromodulatory influences at the neuronal level by controlling neural excitability in small networks, thus not accounting for the influences of neuromodulators at the network level[185].

On the other hand, modulation at the circuitry level has been incorporated into all types of ANNs through meta-learning processes (see Box 3)[92,158,164]. Specifically, in the context of SNNs, research studies have explored the synergy of distinct neuromodulatory signals, such as DA interacting with ACh or 5-HT, through modulated plasticity rules like R-STDP, adjusting plasticity windows based on different neuromodulator releases and yielding efficient, flexible learning[143,154,155]. However, these studies have not considered the influence of these neuromodulators at the cellular level, where they affect the spiking behavior of neurons. Our framework aims to bridge the gap between these two approaches by generating networks with neuronal diversity where neuromodulators act in a neuron-specific manner, incorporating synergy not only among neuromodulators but also between neuromodulators and specific neuronal types. In line with this, a study demonstrated that synaptic credit assignment in SNNs can be guided through cell-type-specific neuromodulation[187]. Similarly, a recent theoretical study modeling IZ neurons has shown that cell-specific neuromodulatory arousal can regulate distinct modes of neural dynamics[25]. Another notable study mitigated catastrophic forgetting in a class-incremental scenario by modulating the population threshold of an SNN[188]. Here, instead of introducing an artificial gating mechanism as with regularized gradient-based algorithms[126], they utilized the inherent gating of SNNs and used neuromodulatory signals to modulate this threshold. Thus, neuronal diversity and neuron-specific neuromodulation appear to be essential components for enhancing learning capabilities in neural networks.

The synergy between cell types and neuromodulation involves complex interactions that fine-tune neural circuit functions, with neuromodulators exerting varying effects based on receptor subtype and neuronal population[19,20,189,190]. For instance, DA can be influenced by the distinct affinities of D1- and D2-type dopamine receptors in different neuronal types, altering the balance between learning from positive and negative reward prediction errors, leading to biased value predictions[191,192]. Similarly, 5-HT selectively reduces synaptic excitation in specific hippocampal interneurons, modulating feedback inhibition in hippocampal pyramidal cells[193].

Moreover, the co-release of neuromodulators and their interactions can further diversify their effects. In the hippocampus, ACh and DA together modulate synaptic plasticity, thereby influencing learning, memory and attention[41,55]. Similarly, 5-HT interacts with DA to influence reward-driven actions and motivation[53,54], while also modulating behavior by opposing the effects of ACh[194,195].

### 6.3. Neuromodulation for robotics

Robots need to maintain homeostatic plasticity when encountering perturbations in real-world environments. Additionally, neuromodulatory mechanism can dynamically adjust synaptic strengths and neuronal excitability to maintain functional stability in the face of ongoing plasticity and environmental fluctuations (see Box 1). Thus, introducing neuromodulatory mechanisms on robots performing online learning will enable them to adapt and widen their operating regime, suitable for goal-directed tasks and closely mimicking biological systems. Furthermore, leveraging heterogeneous populations of cell types would allow networks to switch operation modes dynamically. However, given that robots are edge devices and as such have relatively limited compute, and that neuromodulatory mechanisms add several hyperparameters operating at various spatial and temporal scales, third-factor learning rules that leverage neuromodulation need to be decomposed using low-rank factorization[187].

While no convincing mechanism exists yet, progress has been made in this space: there have been theoretical frameworks that use distributions of neuron parameters (e.g., Lorentzian-distributed spike thresholds) to capture heterogeneity effects on population dynamics, and such models have shown that controlling the degree of heterogeneity adjusts gating, encoding, and decoding properties of SNNs[24]. Additionally, studies leveraging eligibility traces show how neuromodulation-related mechanisms may achieve credit-assignment using a more energy-efficient alternative to backpropagation[91,187], although scalability issues persist. With the advent of large event-based training datasets (e.g., Prophesee Gen1 dataset[196]) and recent advancements focused on closed-loop adaptive neuromorphic systems – where neuromodulation dynamically adjusts synaptic weights and neuron excitability on-chip in real-time[197] – SNNs are becoming more viable for continual learning and robust performance. Thus, these advancements underscore the potential of applying neuromodulated SNNs to robotics, with low-rank models offering a path to mechanisms that replicate neuromodulatory functions – enabling energy-efficient, low-compute neuromorphic robots capable of maintaining robust behavior under environmental disturbances.

## 7. Outlook

We propose a neuromodulatory-aware SNN framework to advance AI systems into continual and adaptable learning paradigms. This biophysically grounded framework leverages neuronal heterogeneity to develop adaptable SNNs that more closely mirror the robustness and flexibility of biological systems. By incorporating learning across diverse spatiotemporal scales, this approach provides a versatile, scalable solution applicable to a wide range of tasks, from behavioral research to complex AI applications.

This framework is well-suited for supervised learning tasks, such as image classification, object recognition, and speech processing[28,91,113–115], where few-shot and multi-task learning are increasingly essential[126,188]. Additionally, the framework's neural heterogeneity enables population-level learning, which offers unique advantages for reward-based tasks[151,198]. Similarly, neuromodulators can be introduced as switches of different modes of neuronal dynamics, through contextual triggers[25]. Thereby enhancing applications in motor control[151] and spatial navigation tasks[154,155], where adaptive and efficient learning is crucial.

Consequently, this neuromodulatory-aware framework seeks to support lifelong RL scenarios, where agents must efficiently and robustly balance short- and long-term rewards and adapt to their environment, much like biological organisms do to optimize survival. Thereby, by introducing different neuromodulatory effects at the microcircuit level, instead of relying on hyperparameter tuning as in abstract RL models[11,50], this framework will potentially explain how neuromodulatory signals reconfigure spiking networks to guide behavior.

Overall, this framework aims to surpass the performance of current ANN models and provide new approaches for efficient and robust adaptable learning, as already demonstrated by recent studies[24,28,57,188]. The introduction of neuronal populations can introduce a technical challenge in determining appropriate proportions; however, it can be addressed through bio-inspired brain ratios or evolutionary algorithms, both enhancing task optimization[26,28].

Despite the successful interpretation of neuromodulatory signals and spiking patterns within this framework, these models are computationally more demanding than perceptron-like neurons (see Figure 1), posing challenges for implementation in large-scale Deep Learning architectures. However, this SNN-based framework can be effectively utilized in neuromorphic hardware, where its low energy consumption and efficiency resulting from asynchronous and sparse activity can be fully exploited, thus leveraging these devices for the next generation of ANNs[8,9,185]. A recent example of this success is the implementation of a memory-heavy *Drosophila* connectome-constrained network on Loihi 2[94,199]. Specifically, by integrating different spiking behaviors in neuromorphic event-driven devices, we can leverage their capacity for asynchronous processing, resulting in more efficient and precise information handling. Similarly, by considering neuromodulatory processes, these devices would potentially be adaptable to changes, thereby allowing for on-chip online learning.

## Acknowledgements

We would like to thank Dr. Gabriel Wainstein for his valuable feedback on the manuscript. This work was supported by a Marie Skłodowska-Curie Global Fellowship Agreement 842,492 to S.R.; Newcastle University Academic Track (NUAcT) Fellowship to S.R.; Fulbright Research Scholarship to S.R.; Lister Institute Prize Fellowship to S.R.; Academy of Medical Sciences Springboard Award to S.R.; a grant from the Air Force Office of Scientific Research (FA9550-23-1-0533) to S.R.; Theoretical Sciences Visiting Program (TSVP) at the Okinawa Institute of Science and Technology (OIST); A.R-G. is supported by a NUAcT PhD studentship. J.M. acknowledges support from the Japan Society for the Promotion of Science (JSPS) and the Japan Science and Technology Agency (JST).

## References

1. Brown, T. B. *et al.* Language Models are Few-Shot Learners. (2020) doi:10.48550/ARXIV.2005.14165.
2. LeCun, Y., Bengio, Y. & Hinton, G. Deep learning. *Nature* **521**, 436–444 (2015).
3. Goodfellow, I., Bengio, Y. & Courville, A. *Deep Learning*. (MIT Press, 2016).
4. Cohen, Y. *et al.* Recent Advances at the Interface of Neuroscience and Artificial Neural Networks. *J. Neurosci.* **42**, 8514–8523 (2022).
5. Hadsell, R., Rao, D., Rusu, A. A. & Pascanu, R. Embracing Change: Continual Learning in Deep Neural Networks. *Trends Cogn. Sci.* **24**, 1028–1040 (2020).
6. Hassabis, D., Kumaran, D., Summerfield, C. & Botvinick, M. Neuroscience-Inspired Artificial Intelligence. *Neuron* **95**, 245–258 (2017).
7. Kudithipudi, D. *et al.* Biological underpinnings for lifelong learning machines. *Nat. Mach. Intell.* **4**, 196–210 (2022).
8. Ivanov, D., Chezhegov, A., Kiselev, M., Grunin, A. & Larionov, D. Neuromorphic artificial intelligence systems. *Front. Neurosci.* **16**, 959626 (2022).
9. Schuman, C. D. *et al.* Opportunities for neuromorphic computing algorithms and applications. *Nat. Comput. Sci.* **2**, 10–19 (2022).
10. Iqbal, A., Mahmood, H., Stuart, G. J., Fishell, G. & Honnuraiah, S. Biologically Realistic Computational Primitives of Neocortex Implemented on Neuromorphic Hardware Improve Vision Transformer Performance. Preprint at https://doi.org/10.1101/2024.10.06.616839 (2024).
11. Doya, K. Metalearning and neuromodulation. *Neural Netw.* **15**, 495–506 (2002).
12. Mei, J., Muller, E. & Ramaswamy, S. Informing deep neural networks by multiscale principles of neuromodulatory systems. *Trends Neurosci.* **45**, 237–250 (2022).
13. Azizzadenesheli, K. *et al.* Neural operators for accelerating scientific simulations and design. *Nat. Rev. Phys.* **6**, 320–328 (2024).
14. Fan, F.-L., Li, Y., Peng, H., Zeng, T. & Wang, F. Towards NeuroAI: Introducing Neuronal Diversity into Artificial Neural Networks. Preprint at https://doi.org/10.48550/ARXIV.2301.09245 (2023).
15. Goulas, A. *et al.* The natural axis of transmitter receptor distribution in the human cerebral cortex. *Proc. Natl. Acad. Sci.* **118**, e2020574118 (2021).
16. Kanari, L. *et al.* Objective Morphological Classification of Neocortical Pyramidal Cells. *Cereb. Cortex* **29**, 1719–1735 (2019).
17. Markram, H. *et al.* Interneurons of the neocortical inhibitory system. *Nat. Rev. Neurosci.* **5**, 793–807 (2004).
18. Zilles, K. & Amunts, K. Anatomical Basis for Functional Specialization. in *fMRI: From Nuclear Spins to Brain Functions* (eds Uludag, K., Ugurbil, K. & Berliner, L.) vol. 30 27–66 (Springer US, Boston, MA, 2015).
19. Marder, E. Neuromodulation of Neuronal Circuits: Back to the Future. *Neuron* **76**, 1–11 (2012).
20. Marder, E. & Thirumalai, V. Cellular, synaptic and network effects of neuromodulation. *Neural Netw.* **15**, 479–493 (2002).
21. Runfeldt, M. J., Sadovsky, A. J. & MacLean, J. N. Acetylcholine functionally reorganizes neocortical microcircuits. *J. Neurophysiol.* **112**, 1205–1216 (2014).
22. Salvan, P. *et al.* Serotonin regulation of behavior via large-scale neuromodulation of serotonin receptor networks. *Nat. Neurosci.* **26**, 53–63 (2023).
23. Shine, J. M. *et al.* Computational models link cellular mechanisms of neuromodulation to large-scale neural dynamics. *Nat. Neurosci.* **24**, 765–776 (2021).
24. Gast, R., Solla, S. A. & Kennedy, A. Neural heterogeneity controls computations in spiking neural networks. *Proc. Natl. Acad. Sci.* **121**, e2311885121 (2024).
25. Munn, B. R. *et al.* Neuronal connected burst cascades bridge macroscale adaptive signatures across arousal states. *Nat. Commun.* **14**, 6846 (2023).
26. Habashy, K. G., Evans, B. D., Goodman, D. F. M. & Bowers, J. S. Adapting to time: why nature evolved a diverse set of neurons. Preprint at https://doi.org/10.48550/ARXIV.2404.14325 (2024).
27. Khona, M. & Fiete, I. R. Attractor and integrator networks in the brain. *Nat. Rev. Neurosci.* **23**, 744–766 (2022).
28. Perez-Nieves, N., Leung, V. C. H., Dragotti, P. L. & Goodman, D. F. M. Neural heterogeneity promotes robust learning. *Nat. Commun.* **12**, 5791 (2021).

29. Stern, M., Istrate, N. & Mazzucato, L. A reservoir of timescales emerges in recurrent circuits with heterogeneous neural assemblies. *eLife* **12**, e86552 (2023).
30. Zheng, H. *et al.* Temporal dendritic heterogeneity incorporated with spiking neural networks for learning multi-timescale dynamics. *Nat. Commun.* **15**, 277 (2024).
31. Tononi, G., Sporns, O. & Edelman, G. M. Measures of degeneracy and redundancy in biological networks. *Proc. Natl. Acad. Sci.* **96**, 3257–3262 (1999).
32. Behera, C. K., Joshi, A., Wang, D.-H., Sharp, T. & Wong-Lin, K. Degeneracy and stability in neural circuits of dopamine and serotonin neuromodulators: A theoretical consideration. *Front. Comput. Neurosci.* **16**, 950489 (2023).
33. Kandel, E. R. & Hawkins, R. D. The Biological Basis of Learning and Individuality. *Sci. Am.* **267**, 78–86 (1992).
34. Wang, L., Zhang, X., Su, H. & Zhu, J. A Comprehensive Survey of Continual Learning: Theory, Method and Application. Preprint at https://doi.org/10.48550/arXiv.2302.00487 (2024).
35. Chen, Z. *Lifelong Machine Learning*. (Springer Nature, Switzerland, 2022).
36. Goodfellow, I. J., Mirza, M., Xiao, D., Courville, A. & Bengio, Y. An Empirical Investigation of Catastrophic Forgetting in Gradient-Based Neural Networks. Preprint at http://arxiv.org/abs/1312.6211 (2015).
37. McCloskey, M. & Cohen, N. J. Catastrophic Interference in Connectionist Networks: The Sequential Learning Problem. in *Psychology of Learning and Motivation* vol. 24 109–165 (Elsevier, 1989).
38. Ratcliff, R. Connectionist models of recognition memory: Constraints imposed by learning and forgetting functions. *Psychol. Rev.* **97**, 285–308 (1990).
39. Dere, E. *et al.* Neuronal histamine and the interplay of memory, reinforcement and emotions. *Behav. Brain Res.* **215**, 209–220 (2010).
40. González, O. C., Sokolov, Y., Krishnan, G. P., Delanois, J. E. & Bazhenov, M. Can sleep protect memories from catastrophic forgetting? *eLife* **9**, e51005 (2020).
41. Brzosko, Z., Mierau, S. B. & Paulsen, O. Neuromodulation of Spike-Timing-Dependent Plasticity: Past, Present, and Future. *Neuron* **103**, 563–581 (2019).
42. Smith, S. J. Chapter 10: Do astrocytes process neural information? in *Progress in Brain Research* vol. 94 119–136 (Elsevier, 1992).
43. Cornell, J., Salinas, S., Huang, H.-Y. & Zhou, M. Microglia regulation of synaptic plasticity and learning and memory. *Neural Regen. Res.* **17**, 705 (2022).
44. Kampa, B. M., Letzkus, J. J. & Stuart, G. J. Dendritic mechanisms controlling spike-timing-dependent synaptic plasticity. *Trends Neurosci.* **30**, 456–463 (2007).
45. Pagkalos, M., Makarov, R. & Poirazi, P. Leveraging dendritic properties to advance machine learning and neuro-inspired computing. *Curr. Opin. Neurobiol.* **85**, 102853 (2024).
46. Toda, T., Parylak, S. L., Linker, S. B. & Gage, F. H. The role of adult hippocampal neurogenesis in brain health and disease. *Mol. Psychiatry* **24**, 67–87 (2019).
47. Schmidgall, S. *et al.* Brain-inspired learning in artificial neural networks: A review. *APL Mach. Learn.* **2**, 021501 (2024).
48. Hasselmo, M. E. *et al.* The Unexplored Territory of Neural Models: Potential Guides for Exploring the Function of Metabotropic Neuromodulation. *Neuroscience* **456**, 143–158 (2021).
49. Hennig, M. H. Ionotropic Receptors Dynamics, Conductance Models. (2022).
50. Doya, K. Modulators of decision making. *Nat. Neurosci.* **11**, 410–416 (2008).
51. Krichmar, J. L. The Neuromodulatory System: A Framework for Survival and Adaptive Behavior in a Challenging World. *Adapt. Behav.* **16**, 385–399 (2008).
52. Benoy, A. & Ramaswamy, S. Histamine in the neocortex: Towards integrating multiscale effectors. *Eur. J. Neurosci.* **60**, 4597–4623 (2024).
53. Boureau, Y.-L. & Dayan, P. Opponency Revisited: Competition and Cooperation Between Dopamine and Serotonin. *Neuropsychopharmacology* **36**, 74–97 (2011).
54. Yagishita, S. Transient and sustained effects of dopamine and serotonin signaling in motivation-related behavior. *Psychiatry Clin. Neurosci.* **74**, 91–98 (2020).
55. Herrero, J. L. *et al.* Acetylcholine contributes through muscarinic receptors to attentional modulation in V1. *Nature* **454**, 1110–1114 (2008).
56. Aston-Jones, G. & Cohen, J. D. An integrative theory of Locus Coeruleus-norepinephrine function: Adaptive Gain and Optimal Performance. *Annu. Rev. Neurosci.* **28**, 403–450 (2005).
57. Munn, B. R., Müller, E. J., Wainstein, G. & Shine, J. M. Coordinated adrenergic and cholinergic neuromodulation facilitate flexible and reliable cortical states that track pupillary fluctuations. (2023).
58. Kar, K. & DiCarlo, J. J. The Quest for an Integrated Set of Neural Mechanisms Underlying Object Recognition in Primates. Preprint at https://doi.org/10.48550/arXiv.2312.05956 (2023).

59. Zador, A. M. A critique of pure learning and what artificial neural networks can learn from animal brains. *Nat. Commun.* **10**, 3770 (2019).
60. Neftci, E. O. & Averbeck, B. B. Reinforcement learning in artificial and biological systems. *Nat. Mach. Intell.* **1**, 133–143 (2019).
61. Manella, L. C., Petersen, N. & Linster, C. Stimulation of the Locus Ceruleus Modulates Signal-to-Noise Ratio in the Olfactory Bulb. *J. Neurosci.* **37**, 11605–11615 (2017).
62. Minces, V., Pinto, L., Dan, Y. & Chiba, A. A. Cholinergic shaping of neural correlations. *Proc. Natl. Acad. Sci.* **114**, 5725–5730 (2017).
63. Chavlis, S. & Poirazi, P. Drawing inspiration from biological dendrites to empower artificial neural networks. *Curr. Opin. Neurobiol.* **70**, 1–10 (2021).
64. Ostojic, S., Brunel, N. & Hakim, V. Synchronization properties of networks of electrically coupled neurons in the presence of noise and heterogeneities. *J. Comput. Neurosci.* **26**, 369–392 (2009).
65. Rich, S., Moradi Chameh, H., Lefebvre, J. & Valiante, T. A. Loss of neuronal heterogeneity in epileptogenic human tissue impairs network resilience to sudden changes in synchrony. *Cell Rep.* **39**, 110863 (2022).
66. Marsat, G. & Maler, L. Neural Heterogeneity and Efficient Population Codes for Communication Signals. *J. Neurophysiol.* **104**, 2543–2555 (2010).
67. Rosenblatt, F. *Principles of Neurodynamics: Perceptrons and the Theory of Brain Mechanisms*. vol. 55 (Spartan books Washington, DC, 1962).
68. Kar, K., Kubilius, J., Schmidt, K., Issa, E. B. & DiCarlo, J. J. Evidence that recurrent circuits are critical to the ventral stream's execution of core object recognition behavior. *Nat. Neurosci.* **22**, 974–983 (2019).
69. Banino, A. *et al.* Vector-based navigation using grid-like representations in artificial agents. *Nature* **557**, 429–433 (2018).
70. Bertens, P. & Lee, S.-W. Network of evolvable neural units can learn synaptic learning rules and spiking dynamics. *Nat. Mach. Intell.* **2**, 791–799 (2020).
71. Mei, J., Meshkinnejad, R. & Mohsenzadeh, Y. Effects of neuromodulation-inspired mechanisms on the performance of deep neural networks in a spatial learning task. *iScience* **26**, 106026 (2023).
72. Smith, C. M., Thompson-Schill, S. L. & Schapiro, A. C. *Rapid Learning of Temporal Dependencies at Multiple Timescales*. http://biorxiv.org/lookup/doi/10.1101/2024.01.15.575748 (2024) doi:10.1101/2024.01.15.575748.
73. Dayan, P. & Abbott, L. F. *Theoretical Neuroscience: Computational and Mathematical Modeling of Neural Systems*. (Massachusetts Institute of Technology Press, Cambridge, Mass, 2001).
74. Kao, J. C. Considerations in using recurrent neural networks to probe neural dynamics. *J. Neurophysiol.* **122**, 2504–2521 (2019).
75. Legenstein, R., Chase, S. M., Schwartz, A. B. & Maass, W. A Reward-Modulated Hebbian Learning Rule Can Explain Experimentally Observed Network Reorganization in a Brain Control Task. *J. Neurosci.* **30**, 8400–8410 (2010).
76. Song, H. F., Yang, G. R. & Wang, X.-J. Training Excitatory-Inhibitory Recurrent Neural Networks for Cognitive Tasks: A Simple and Flexible Framework. *PLOS Comput. Biol.* **12**, e1004792 (2016).
77. Greedy, W., Zhu, H. W., Pemberton, J., Mellor, J. & Costa, R. P. Single-phase deep learning in cortico-cortical networks. Preprint at http://arxiv.org/abs/2206.11769 (2022).
78. Sacramento, J., Costa, R. P., Bengio, Y. & Senn, W. Dendritic cortical microcircuits approximate the backpropagation algorithm. Preprint at http://arxiv.org/abs/1810.11393 (2018).
79. Izhikevich, E. M. Which Model to Use for Cortical Spiking Neurons? *IEEE Trans. Neural Netw.* **15**, 1063–1070 (2004).
80. Izhikevich, E. M. Hybrid spiking models. *Philos. Trans. R. Soc. Math. Phys. Eng. Sci.* **368**, 5061–5070 (2010).
81. Hodgkin, A. L. & Huxley, A. F. A quantitative description of membrane current and its application to conduction and excitation in nerve. *J. Physiol.* **117**, 500–544 (1952).
82. Billeh, Y. N. *et al.* Systematic Integration of Structural and Functional Data into Multi-scale Models of Mouse Primary Visual Cortex. *Neuron* **106**, 388-403.e18 (2020).
83. Gouwens, N. W. *et al.* Systematic generation of biophysically detailed models for diverse cortical neuron types. *Nat. Commun.* **9**, 710 (2018).
84. Halnes, G., Augustinaite, S., Heggelund, P., Einevoll, G. T. & Migliore, M. A Multi-Compartment Model for Interneurons in the Dorsal Lateral Geniculate Nucleus. *PLOS Comput. Biol.* **7**, e1002160 (2011).
85. Tzilivaki, A., Kastellakis, G. & Poirazi, P. Challenging the point neuron dogma: FS basket cells as 2-stage nonlinear integrators. *Nat. Commun.* **10**, 3664 (2019).
86. Carnevale, N. T. & Hines, M. L. *The NEURON Book*. (Cambridge University Press, 2006).
87. Hines, M., Davison, A. & Muller, E. NEURON and Python. *Front. Neuroinformatics* **3**, (2009).
88. Stimberg, M., Brette, R. & Goodman, D. F. Brian 2, an intuitive and efficient neural simulator. *eLife* **8**, e47314 (2019).

89. Pagkalos, M., Chavlis, S. & Poirazi, P. Introducing the Dendrify framework for incorporating dendrites to spiking neural networks. *Nat. Commun.* **14**, 131 (2023).
90. Abbott, L. F. Lapicque's introduction of the integrate-and-fire model neuron (1907). *Brain Res. Bull.* **50**, 303–304 (1999).
91. Bellec, G. *et al.* A solution to the learning dilemma for recurrent networks of spiking neurons. *Nat. Commun.* **11**, 3625 (2020).
92. Schmidgall, S. & Hays, J. Meta-SpikePropamine: learning to learn with synaptic plasticity in spiking neural networks. *Front. Neurosci.* **17**, 1183321 (2023).
93. Zhu, R.-J., Zhao, Q., Li, G. & Eshraghian, J. K. SpikeGPT: Generative Pre-trained Language Model with Spiking Neural Networks. Preprint at http://arxiv.org/abs/2302.13939 (2023).
94. Orchard, G. *et al.* Efficient Neuromorphic Signal Processing with Loihi 2. Preprint at https://doi.org/10.48550/arXiv.2111.03746 (2021).
95. Furber, S. B., Galluppi, F., Temple, S. & Plana, L. A. The SpiNNaker Project. *Proc. IEEE* **102**, 652–665 (2014).
96. Fang, W. *et al.* Incorporating Learnable Membrane Time Constant to Enhance Learning of Spiking Neural Networks. in *2021 IEEE/CVF International Conference on Computer Vision (ICCV)* 2641–2651 (IEEE, Montreal, QC, Canada, 2021). doi:10.1109/ICCV48922.2021.00266.
97. Ivanov, V. A. & Michmizos, K. P. Increasing Liquid State Machine Performance with Edge-of-Chaos Dynamics Organized by Astrocyte-modulated Plasticity. Preprint at http://arxiv.org/abs/2111.01760 (2021).
98. Teeter, C. *et al.* Generalized leaky integrate-and-fire models classify multiple neuron types. *Nat. Commun.* **9**, 709 (2018).
99. Izhikevich, E. M. Simple model of spiking neurons. *IEEE Trans. Neural Netw.* **14**, 1569–1572 (2003).
100. Rumelhart, D. E., Hintont, G. E. & Williams, R. J. Learning representations by back-propagating errors. (1986).
101. Daube, C. *et al.* Grounding deep neural network predictions of human categorization behavior in understandable functional features: The case of face identity. *Patterns* **2**, 100348 (2021).
102. Jacob, G., Pramod, R. T., Katti, H. & Arun, S. P. Qualitative similarities and differences in visual object representations between brains and deep networks. *Nat. Commun.* **12**, 1872 (2021).
103. Ratan Murty, N. A., Bashivan, P., Abate, A., DiCarlo, J. J. & Kanwisher, N. Computational models of category-selective brain regions enable high-throughput tests of selectivity. *Nat. Commun.* **12**, 5540 (2021).
104. Xu, Y. & Vaziri-Pashkam, M. Limits to visual representational correspondence between convolutional neural networks and the human brain. *Nat. Commun.* **12**, 2065 (2021).
105. Lillicrap, T. P., Santoro, A., Marris, L., Akerman, C. J. & Hinton, G. Backpropagation and the brain. *Nat. Rev. Neurosci.* **21**, 335–346 (2020).
106. Marblestone, A. H., Wayne, G. & Kording, K. P. Toward an Integration of Deep Learning and Neuroscience. *Front. Comput. Neurosci.* **10**, (2016).
107. Richards, B. A. *et al.* A deep learning framework for neuroscience. *Nat. Neurosci.* **22**, 1761–1770 (2019).
108. Whittington, J. C. R. & Bogacz, R. Theories of Error Back-Propagation in the Brain. *Trends Cogn. Sci.* **23**, 235–250 (2019).
109. Sussillo, D. & Abbott, L. F. Generating Coherent Patterns of Activity from Chaotic Neural Networks. *Neuron* **63**, 544–557 (2009).
110. DePasquale, B., Cueva, C. J., Rajan, K., Escola, G. S. & Abbott, L. F. full-FORCE: A target-based method for training recurrent networks. *PLOS ONE* **13**, e0191527 (2018).
111. Nøkland, A. Direct Feedback Alignment Provides Learning in Deep Neural Networks. (2016).
112. Lillicrap, T. P., Cownden, D., Tweed, D. B. & Akerman, C. J. Random synaptic feedback weights support error backpropagation for deep learning. *Nat. Commun.* **7**, 13276 (2016).
113. Zenke, F. & Vogels, T. P. The Remarkable Robustness of Surrogate Gradient Learning for Instilling Complex Function in Spiking Neural Networks. *Neural Comput.* **33**, 899–925 (2021).
114. Neftci, E. O., Mostafa, H. & Zenke, F. Surrogate Gradient Learning in Spiking Neural Networks: Bringing the Power of Gradient-Based Optimization to Spiking Neural Networks. *IEEE Signal Process. Mag.* **36**, 51–63 (2019).
115. Scherr, F., Stöckl, C. & Maass, W. *One-Shot Learning with Spiking Neural Networks*. http://biorxiv.org/lookup/doi/10.1101/2020.06.17.156513 (2020) doi:10.1101/2020.06.17.156513.
116. Baldi, P. & Sadowski, P. J. Understanding Dropout. (2013).
117. Zeng, G., Chen, Y., Cui, B. & Yu, S. Continual learning of context-dependent processing in neural networks. *Nat. Mach. Intell.* **1**, 364–372 (2019).
118. Aljundi, R., Babiloni, F., Elhoseiny, M., Rohrbach, M. & Tuytelaars, T. Memory Aware Synapses: Learning what (not) to forget. Preprint at http://arxiv.org/abs/1711.09601 (2018).

119. Kolouri, S., Ketz, N. A., Pilly, P. K. & Soltoggio, A. Sliced Cramér synaptic consolidation for preserving deeply learned representations. (2020).
120. Zenke, F., Poole, B. & Ganguli, S. Continual Learning Through Synaptic Intelligence. (2017).
121. Kirkpatrick, J. *et al.* Overcoming catastrophic forgetting in neural networks. *Proc. Natl. Acad. Sci.* **114**, 3521–3526 (2017).
122. Acharya, J. *et al.* Dendritic Computing: Branching Deeper into Machine Learning. *Neuroscience* **489**, 275–289 (2022).
123. Bono, J. & Clopath, C. Modeling somatic and dendritic spike mediated plasticity at the single neuron and network level. *Nat. Commun.* **8**, 706 (2017).
124. Kastellakis, G., Silva, A. J. & Poirazi, P. Linking Memories across Time via Neuronal and Dendritic Overlaps in Model Neurons with Active Dendrites. *Cell Rep.* **17**, 1491–1504 (2016).
125. Limbacher, T. & Legenstein, R. Emergence of Stable Synaptic Clusters on Dendrites Through Synaptic Rewiring. *Front. Comput. Neurosci.* **14**, 57 (2020).
126. van de Ven, G. M. & Tolias, A. S. Three scenarios for continual learning. Preprint at http://arxiv.org/abs/1904.07734 (2019).
127. Löwel, S. & Singer, W. Selection of Intrinsic Horizontal Connections in the Visual Cortex by Correlated Neuronal Activity. *Science* **255**, 209–212 (1992).
128. Hwu, T. & Krichmar, J. L. A neural model of schemas and memory encoding. *Biol. Cybern.* **114**, 169–186 (2020).
129. Markram, H. A history of spike-timing-dependent plasticity. *Front. Synaptic Neurosci.* **3**, (2011).
130. Markram, H., Lübke, J., Frotscher, M. & Sakmann, B. Regulation of Synaptic Efficacy by Coincidence of Postsynaptic APs and EPSPs. *Science* **275**, 213–215 (1997).
131. Triche, A., Maida, A. S. & Kumar, A. Exploration in neo-Hebbian reinforcement learning: Computational approaches to the exploration–exploitation balance with bio-inspired neural networks. *Neural Netw.* **151**, 16–33 (2022).
132. Lazar, A. SORN: a Self-organizing Recurrent Neural Network. *Front. Comput. Neurosci.* **3**, (2009).
133. Manna, D. L., Vicente-Sola, A., Kirkland, P., Bihl, T. J. & Di Caterina, G. Simple and complex spiking neurons: perspectives and analysis in a simple STDP scenario. *Neuromorphic Comput. Eng.* **2**, 044009 (2022).
134. Saranirad, V., McGinnity, T. M., Dora, S. & Coyle, D. DoB-SNN: A New Neuron Assembly-Inspired Spiking Neural Network for Pattern Classification. in *2021 International Joint Conference on Neural Networks (IJCNN)* 1–6 (IEEE, Shenzhen, China, 2021). doi:10.1109/IJCNN52387.2021.9534283.
135. Gjorgjieva, J., Clopath, C., Audet, J. & Pfister, J.-P. A triplet spike-timing–dependent plasticity model generalizes the Bienenstock–Cooper–Munro rule to higher-order spatiotemporal correlations. *Proc. Natl. Acad. Sci.* **108**, 19383–19388 (2011).
136. Naud, R. & Sprekeler, H. Sparse bursts optimize information transmission in a multiplexed neural code. *Proc. Natl. Acad. Sci.* **115**, (2018).
137. Payeur, A., Guerguiev, J., Zenke, F., Richards, B. A. & Naud, R. Burst-dependent synaptic plasticity can coordinate learning in hierarchical circuits. *Nat. Neurosci.* **24**, 1010–1019 (2021).
138. Lisman, J., Grace, A. A. & Duzel, E. A neoHebbian framework for episodic memory; role of dopamine-dependent late LTP. *Trends Neurosci.* **34**, 536–547 (2011).
139. McFarlan, A. R. *et al.* The plasticitome of cortical interneurons. *Nat. Rev. Neurosci.* **24**, 80–97 (2023).
140. Dora, S., Bohte, S. M. & Pennartz, C. M. A. Deep Gated Hebbian Predictive Coding Accounts for Emergence of Complex Neural Response Properties Along the Visual Cortical Hierarchy. *Front. Comput. Neurosci.* **15**, 666131 (2021).
141. Soltoggio, A. Short-term plasticity as cause–effect hypothesis testing in distal reward learning. *Biol. Cybern.* **109**, 75–94 (2015).
142. Miconi, T. Biologically plausible learning in recurrent neural networks reproduces neural dynamics observed during cognitive tasks. *eLife* **6**, e20899 (2017).
143. Pedrosa, V. & Clopath, C. The Role of Neuromodulators in Cortical Plasticity. A Computational Perspective. *Front. Synaptic Neurosci.* **8**, (2017).
144. Zhang, J.-C., Lau, P.-M. & Bi, G.-Q. Gain in sensitivity and loss in temporal contrast of STDP by dopaminergic modulation at hippocampal synapses. *Proc. Natl. Acad. Sci.* **106**, 13028–13033 (2009).
145. Frémaux, N. & Gerstner, W. Neuromodulated Spike-Timing-Dependent Plasticity, and Theory of Three-Factor Learning Rules. *Front. Neural Circuits* **9**, (2016).
146. Izhikevich, E. M. Solving the Distal Reward Problem through Linkage of STDP and Dopamine Signaling. *Cereb. Cortex* **17**, 2443–2452 (2007).
147. Mozafari, M., Kheradpisheh, S. R., Masquelier, T., Nowzari-Dalini, A. & Ganjtabesh, M. First-Spike-Based Visual Categorization Using Reward-Modulated STDP. *IEEE Trans. Neural Netw. Learn. Syst.* **29**, 6178–6190 (2018).

148. Szatmáry, B. & Izhikevich, E. M. Spike-Timing Theory of Working Memory. *PLoS Comput. Biol.* **6**, e1000879 (2010).
149. Frémaux, N., Sprekeler, H. & Gerstner, W. Reinforcement Learning Using a Continuous Time Actor-Critic Framework with Spiking Neurons. *PLoS Comput. Biol.* **9**, e1003024 (2013).
150. Nazari, S. *A New Unsupervised/Reinforcement Learning Method In Spiking Pattern Classification Networks*. https://www.researchsquare.com/article/rs-3560563/v1 (2023) doi:10.21203/rs.3.rs-3560563/v1.
151. Chung, S. & Kozma, R. Reinforcement Learning with Feedback-modulated TD-STDP. Preprint at http://arxiv.org/abs/2008.13044 (2020).
152. Vasilaki, E., Frémaux, N., Urbanczik, R., Senn, W. & Gerstner, W. Spike-Based Reinforcement Learning in Continuous State and Action Space: When Policy Gradient Methods Fail. *PLoS Comput. Biol.* **5**, e1000586 (2009).
153. Golden, R., Rossa, M. & Olayinka, T. Parametrization of Neuromodulation in Reinforcement Learning. (2016).
154. Zannone, S., Brzosko, Z., Paulsen, O. & Clopath, C. Acetylcholine-modulated plasticity in reward-driven navigation: a computational study. *Sci. Rep.* **8**, 9486 (2018).
155. Wert-Carvajal, C., Reneaux, M., Tchumatchenko, T. & Clopath, C. Dopamine and serotonin interplay for valence-based spatial learning. *Cell Rep.* **39**, 110645 (2022).
156. Fogel, D. B. & Fogel, L. J. An introduction to evolutionary programming. in *Artificial Evolution* (eds Alliot, J.-M., Lutton, E., Ronald, E., Schoenauer, M. & Snyers, D.) vol. 1063 21–33 (Springer Berlin Heidelberg, Berlin, Heidelberg, 1996).
157. Koza, JohnR. Genetic programming as a means for programming computers by natural selection. *Stat. Comput.* **4**, (1994).
158. Jordan, J., Schmidt, M., Senn, W. & Petrovici, M. A. Evolving interpretable plasticity for spiking networks. *eLife* **10**, e66273 (2021).
159. Velez, R. & Clune, J. Diffusion-based neuromodulation can eliminate catastrophic forgetting in simple neural networks. *PLOS ONE* **12**, e0187736 (2017).
160. Sutton, R. S. & Barto, A. G. *Reinforcement Learning: An Introduction*. (The MIT Press, Cambridge, Massachusetts, 2018).
161. Schultz, W. Dopamine reward prediction-error signalling: a two-component response. *Nat. Rev. Neurosci.* **17**, 183–195 (2016).
162. Schultz, W. Predictive Reward Signal of Dopamine Neurons. *J. Neurophysiol.* **80**, 1–27 (1998).
163. Botvinick, M. *et al.* Reinforcement Learning, Fast and Slow. *Trends Cogn. Sci.* **23**, 408–422 (2019).
164. Miconi, T. Learning to acquire novel cognitive tasks with evolution, plasticity and meta-meta-learning. Preprint at http://arxiv.org/abs/2112.08588 (2023).
165. Vecoven, N., Ernst, D., Wehenkel, A. & Drion, G. Introducing neuromodulation in deep neural networks to learn adaptive behaviours. *PLOS ONE* **15**, e0227922 (2020).
166. Major, G., Larkum, M. E. & Schiller, J. Active Properties of Neocortical Pyramidal Neuron Dendrites. *Annu. Rev. Neurosci.* **36**, 1–24 (2013).
167. Wu, X., Liu, X., Li, W. & Wu, Q. Improved Expressivity Through Dendritic Neural Networks. (2018).
168. Häusser, M. & Mel, B. Dendrites: bug or feature? *Curr. Opin. Neurobiol.* **13**, 372–383 (2003).
169. Poirazi, P., Brannon, T. & Mel, B. W. Pyramidal Neuron as Two-Layer Neural Network. *Neuron* **37**, 989–999 (2003).
170. Iyer, A. *et al.* Avoiding Catastrophe: Active Dendrites Enable Multi-Task Learning in Dynamic Environments. *Front. Neurorobotics* **16**, 846219 (2022).
171. Wybo, W. A. M. *et al.* NMDA-driven dendritic modulation enables multitask representation learning in hierarchical sensory processing pathways. *Proc. Natl. Acad. Sci.* **120**, e2300558120 (2023).
172. Ward, M. & Rhodes, O. Beyond LIF Neurons on Neuromorphic Hardware. *Front. Neurosci.* **16**, 881598 (2022).
173. Rusu, A. A. *et al.* Progressive Neural Networks. Preprint at http://arxiv.org/abs/1606.04671 (2022).
174. Parisi, G. I., Kemker, R., Part, J. L., Kanan, C. & Wermter, S. Continual lifelong learning with neural networks: A review. *Neural Netw.* **113**, 54–71 (2019).
175. Dale, H. Pharmacology and Nerve-Endings. (1934).
176. Tritsch, N. X., Granger, A. J. & Sabatini, B. L. Mechanisms and functions of GABA co-release. *Nat. Rev. Neurosci.* **17**, 139–145 (2016).
177. Vaaga, C. E., Borisovska, M. & Westbrook, G. L. Dual-transmitter neurons: Functional implications of co-release and co-transmission. *Curr. Opin. Neurobiol.* **0**, 25–32 (2014).
178. Cornford, J. *et al. Learning to Live with Dale's Principle: ANNs with Separate Excitatory and Inhibitory Units*. http://biorxiv.org/lookup/doi/10.1101/2020.11.02.364968 (2020) doi:10.1101/2020.11.02.364968.
179. Li, P., Cornford, J., Ghosh, A. & Richards, B. *Learning Better with Dale's Law: A Spectral Perspective*. http://biorxiv.org/lookup/doi/10.1101/2023.06.28.546924 (2023) doi:10.1101/2023.06.28.546924.

180. Amit, D. J., Campbell, C. & Wong, K. Y. M. The interaction space of neural networks with sign-constrained synapses. *J. Phys. Math. Gen.* **22**, 4687–4693 (1989).
181. Murphy-Royal, C., Ching, S. & Papouin, T. A conceptual framework for astrocyte function. *Nat. Neurosci.* **26**, 1848–1856 (2023).
182. Kozachkov, L., Kastanenka, K. V. & Krotov, D. Building transformers from neurons and astrocytes. *Proc. Natl. Acad. Sci.* **120**, e2219150120 (2023).
183. Vaswani, A. *et al.* Attention is All you Need. (2017).
184. Van Schaik, A. *et al.* A log-domain implementation of the Mihalas-Niebur neuron model. in *Proceedings of 2010 IEEE International Symposium on Circuits and Systems* 4249–4252 (IEEE, Paris, France, 2010). doi:10.1109/ISCAS.2010.5537563.
185. Ribar, L. & Sepulchre, R. Neuromorphic Control: Designing Multiscale Mixed-Feedback Systems. *IEEE Control Syst.* **41**, 34–63 (2021).
186. Sepulchre, R. Spiking Control Systems. Preprint at http://arxiv.org/abs/2112.03565 (2022).
187. Liu, Y. H., Smith, S., Mihalas, S., Shea-Brown, E. & Sümbül, U. Cell-type–specific neuromodulation guides synaptic credit assignment in a spiking neural network. *Proc. Natl. Acad. Sci.* **118**, e2111821118 (2021).
188. Hammouamri, I., Masquelier, T. & Wilson, D. Mitigating Catastrophic Forgetting in Spiking Neural Net- works through Threshold Modulation. (2023).
189. Briand, L. A., Gritton, H., Howe, W. M., Young, D. A. & Sarter, M. Modulators in concert for cognition: Modulator interactions in the prefrontal cortex. *Prog. Neurobiol.* **83**, 69–91 (2007).
190. Hansen, J. Y. *et al.* Mapping neurotransmitter systems to the structural and functional organization of the human neocortex. *Nat. Neurosci.* **25**, 1569–1581 (2022).
191. Romero Pinto, S. & Uchida, N. *Tonic Dopamine and Biases in Value Learning Linked through a Biologically Inspired Reinforcement Learning Model*. http://biorxiv.org/lookup/doi/10.1101/2023.11.10.566580 (2023) doi:10.1101/2023.11.10.566580.
192. Vijayraghavan, S., Wang, M., Birnbaum, S. G., Williams, G. V. & Arnsten, A. F. T. Inverted-U dopamine D1 receptor actions on prefrontal neurons engaged in working memory. *Nat. Neurosci.* **10**, 376–384 (2007).
193. Winterer, J. *et al.* Cell-Type-Specific Modulation of Feedback Inhibition by Serotonin in the Hippocampus. *J. Neurosci.* **31**, 8464–8475 (2011).
194. Sparks, D. W. *et al.* Opposing Cholinergic and Serotonergic Modulation of Layer 6 in Prefrontal Cortex. *Front. Neural Circuits* **11**, 107 (2018).
195. Steckler, T. & Sahgal, A. The role of serotonergic-cholinergic interactions in the mediation of cognitive behaviour. *Behav. Brain Res.* **67**, 165–199 (1995).
196. de Tournemire, P., Nitti, D., Perot, E., Migliore, D. & Sironi, A. A Large Scale Event-based Detection Dataset for Automotive. Preprint at https://doi.org/10.48550/ARXIV.2001.08499 (2020).
197. Contreras, L. F. H. *et al.* Neuromorphic Neuromodulation: Towards the next generation of on-device AI-revolution in electroceuticals. Preprint at https://doi.org/10.48550/arXiv.2307.12471 (2023).
198. Haşegan, D. *et al.* Training spiking neuronal networks to perform motor control using reinforcement and evolutionary learning. *Front. Comput. Neurosci.* **16**, 1017284 (2022).
199. Wang, F. *et al.* Neuromorphic Simulation of Drosophila Melanogaster Brain Connectome on Loihi 2. Preprint at https://doi.org/10.48550/arXiv.2508.16792 (2025).
200. Stewart, K., Orchard, G., Shrestha, S. B. & Neftci, E. Online Few-shot Gesture Learning on a Neuromorphic Processor. Preprint at http://arxiv.org/abs/2008.01151 (2020).
201. Schmetterling, R., Forni, F., Franci, A. & Sepulchre, R. Neuromorphic Control of a Pendulum. Preprint at http://arxiv.org/abs/2404.05339 (2024).
202. McCulloch, W. S. & Pitts, W. A logical calculus of the ideas immanent in nervous activity. *Bull. Math. Biophys.* **5**, 115–133 (1943).
203. Cho, K. *et al.* Learning Phrase Representations using RNN Encoder-Decoder for Statistical Machine Translation. Preprint at https://doi.org/10.48550/ARXIV.1406.1078 (2014).
204. Hochreiter, S. & Schmidhuber, J. Long Short-Term Memory. *Neural Comput.* **9**, 1735–1780 (1997).
205. Ermentrout, G. B. & Kopell, N. Parabolic Bursting in an Excitable System Coupled with a Slow Oscillation. *SIAM J. Appl. Math.* **46**, 233–253 (1986).
206. Badel, L. *et al.* Dynamic I-V Curves Are Reliable Predictors of Naturalistic Pyramidal-Neuron Voltage Traces. *J. Neurophysiol.* **99**, 656–666 (2008).
207. Izhikevich, E. M. Resonate-and-fire neurons. *Neural Netw.* **14**, 883–894 (2001).

208. Smith, G. D., Cox, C. L., Sherman, S. M. & Rinzel, J. Fourier Analysis of Sinusoidally Driven Thalamocortical Relay Neurons and a Minimal Integrate-and-Fire-or-Burst Model. *J. Neurophysiol.* **83**, 588–610 (2000).
209. FitzHugh, R. Impulses and Physiological States in Theoretical Models of Nerve Membrane. *Biophys. J.* **1**, 445–466 (1961).
210. Gerstner, W. Spike-response model. *Scholarpedia* **3**, 1343 (2008).
211. Morris, C. & Lecar, H. Voltage oscillations in the barnacle giant muscle fiber. *Biophys. J.* **35**, 193–213 (1981).
212. Brette, R. & Gerstner, W. Adaptive Exponential Integrate-and-Fire Model as an Effective Description of Neuronal Activity. *J. Neurophysiol.* **94**, 3637–3642 (2005).
213. Rose, R. & Hindmarsh, J. The assembly of ionic currents in a thalamic neuron I. The three-dimensional model. *Proc. R. Soc. Lond. B Biol. Sci.* **237**, 267–288 (1989).
214. Wilson, H. R. Simplified Dynamics of Human and Mammalian Neocortical Neurons. *J. Theor. Biol.* **200**, 375–388 (1999).
215. Zemansky, M. W. & Dittman, R. H. *Heat and Thermodynamics*. (The McGraw-Hill Companies, Inc., Boston, MA, 1997).
216. Izhikevich, E. M. Polychronization: Computation with Spikes. *Neural Comput.* **18**, 245–282 (2006).
217. Mihalaş, Ş. & Niebur, E. A Generalized Linear Integrate-and-Fire Neural Model Produces Diverse Spiking Behaviors. *Neural Comput.* **21**, 704–718 (2009).
218. Beaulieu, S. *et al.* Learning to Continually Learn. Preprint at http://arxiv.org/abs/2002.09571 (2020).
219. Wilson, D. G., Cussat-Blanc, S., Luga, H. & Harrington, K. Neuromodulated Learning in Deep Neural Networks. Preprint at http://arxiv.org/abs/1812.03365 (2018).